\documentclass[12pt,fleqn]{article}
\usepackage{english,latexsym,amssymb,graphics}
\usepackage{epsfig,feynmp}

\makeatletter
\newif\if@preliminary
\@preliminaryfalse
\def\preliminary{\@preliminarytrue}
\def\preprintno#1{\def\@preprintno{#1}}
\def\address#1{\def\@address{\textit{#1}}}

\def\abstract#1{\def\@abstract{#1}}
\newlength\preprintnoskip
\newlength\abstractwidth
\@titlepagetrue
\renewcommand\maketitle{\begin{titlepage}%
  \let\footnotesize\small
  \hfill\parbox{\preprintnoskip}{%
  \begin{flushright}\@preprintno\end{flushright}}\hspace*{1cm}
  \vskip 60\p@
  \begin{center}%
    {\Large\bf\boldmath \@title \par}\vskip 1cm%
    {\sc\@author \par}\vskip 3mm%
    {\@address \par}%
    \if@preliminary
      \vskip 2cm {\large\sf PRELIMINARY DRAFT \par \@date}%
    \fi
  \end{center}\par
  \@thanks
  \vfill
  \begin{center}%
    \parbox{\abstractwidth}{\centerline{\bf\abstractname}%
    \vskip 3mm%
    \@abstract}
  \end{center}
  \end{titlepage}%
  \setcounter{footnote}{0}%
  \let\thanks\relax\let\maketitle\relax
  \gdef\@thanks{}\gdef\@author{}\gdef\@address{}%
  \gdef\@title{}\gdef\@abstract{}\gdef\@preprintno{}
}%
\def\@citex[#1]#2{\if@filesw\immediate\write\@auxout{\string\citation{#2}}\fi
  \def\@citea{}\@cite{\@for\@citeb:=#2\do
    {\@citea\def\@citea{,\penalty\@m}\@ifundefined
       {b@\@citeb}{{\bf ?}\@warning
       {Citation `\@citeb' on page \thepage \space undefined}}%
\hbox{\csname b@\@citeb\endcsname}}}{#1}}
\def\citerange{\@ifnextchar [{\@tempswatrue\@citexr}{\@tempswafalse\@citexr[]}}
\def\@citexr[#1]#2{\if@filesw\immediate\write\@auxout{\string\citation{#2}}\fi
  \def\@citea{}\@cite{\@for\@citeb:=#2\do
    {\@citea\def\@citea{--\penalty\@m}\@ifundefined
       {b@\@citeb}{{\bf ?}\@warning
       {Citation `\@citeb' on page \thepage \space undefined}}%
\hbox{\csname b@\@citeb\endcsname}}}{#1}}
\long\def\@makecaption#1#2{%
  \vskip\abovecaptionskip
  \sbox\@tempboxa{#1: \emph{#2}}%
  \ifdim \wd\@tempboxa >\hsize
    #1: \emph{#2}\par
  \else
    \hbox to\hsize{\hfil\box\@tempboxa\hfil}%
  \fi
  \vskip\belowcaptionskip}
\newcommand{\beq}{\begin{eqnarray}}
\newcommand{\eeq}{\end{eqnarray}}
\newcommand{\nn}{\noindent}
\newcommand{\non}{\nonumber}
\newcommand{\str}{\vphantom{\bigg(}}
\newcommand{\pskip}{\vspace{\baselineskip}}
\newcommand{\s}{\\ \vspace*{-3.5mm} } 
\newcommand{\ee}{$e^+e^-$}

\newcommand{\lra}{\longrightarrow}
\makeatletter
\def\fmslash{\@ifnextchar[{\fmsl@sh}{\fmsl@sh[0mu]}}
\def\fmsl@sh[#1]#2{%
  \mathchoice
    {\@fmsl@sh\displaystyle{#1}{#2}}%
    {\@fmsl@sh\textstyle{#1}{#2}}%
    {\@fmsl@sh\scriptstyle{#1}{#2}}%
    {\@fmsl@sh\scriptscriptstyle{#1}{#2}}}
\def\@fmsl@sh#1#2#3{\m@th\ooalign{$\hfil#1\mkern#2/\hfil$\crcr$#1#3$}}
\makeatother
\def\fmfL(#1,#2,#3)#4{\put(#1,#2){\makebox(0,0)[#3]{#4}}}

\begin{document}
\baselineskip16pt   

\preprintno{%
hep-ph/9903229\\
DESY 99/001\\
TTP99-02\\
PM/99-01}

\title{%
 Testing Higgs Self-couplings at \ee ~Linear Colliders
}
\author{%
 A.~Djouadi$^1$, W.~Kilian$^2$, M.~Muhlleitner$^3$ and 
 P.M.~Zerwas$^3$ 
}
\address{%
  $^1$Lab. de Physique Math\'{e}matique, 
        Universit\'{e} Montpellier, F-34095 Montpellier Cedex 5\\
  $^2$Institut f\"ur Theoretische Teilchenphysik, 
        Universit\"at Karlsruhe, D-76128 Karlsruhe\\
  $^3$Deutsches Elektronensynchrotron DESY, D-22603 Hamburg
}
\abstract{%
  To establish the Higgs mechanism {\it sui generis} experimentally,
  the self-energy potential of the Higgs field must be
  recon\-struc\-ted. This task requires the measurement of the
  trilinear and quadrilinear self-couplings, as predicted, for
  instance, in the Standard Model or in supersymmetric theories. The
  couplings can be probed in multiple Higgs production at
  high-luminosity \ee ~linear colliders.  Complementing earlier
  studies to develop a coherent picture of the trilinear couplings, we
  have analyzed the production of pairs of neutral Higgs bosons in all
  relevant channels of double Higgs-strahlung, associated multiple
  Higgs production and $WW$/$ZZ$ fusion to Higgs pairs. }
\maketitle
\subsection*{1. Introduction}

{\bf 1.}  The Higgs mechanism \cite{higgs} is a cornerstone in the
electroweak sector of the Standard Model (SM) \cite{gunion}. The
electroweak gauge bosons and the fundamental matter particles acquire
masses through the interaction with a scalar field. The
self-interaction of the scalar field leads to a non-zero field
strength $v=(\sqrt{2} G_F)^{-1/2} = 246$~GeV in the ground state,
inducing the spontaneous breaking of the electroweak ${\rm
  SU(2)_L\times U(1)_Y}$ symmetry down to the electromagnetic ${\rm
  U(1)_{EM}}$ symmetry.\s

To establish the Higgs mechanism {\it sui generis} experimentally, the
characteristic self-energy potential of the Standard Model,
\beq 
V = \lambda \left[|\varphi|^2 -\textstyle{\frac{1}{2}} v^2 \right]^2 
\eeq 
with a minimum at $\langle \varphi \rangle_0 = v/\sqrt{2}$, must be
reconstructed once the Higgs particle will have been discovered. This
experimental task requires the measurement of the trilinear and
quadrilinear self-couplings of the Higgs boson. The self-couplings are
uniquely determined in the Standard Model by the mass of the Higgs
boson which is related to the quadrilinear coupling $\lambda$ by $M_H =
\sqrt{2\lambda} v$. Introducing the physical Higgs field $H$,
\beq
\varphi = \frac{1}{\sqrt{2}}\left( \begin{array}{c} 0 \\ v+H 
\end{array} \right)
\eeq
the trilinear vertex of the Higgs field $H$ is given by the coefficient 
\beq
\lambda_{HHH} = 3 M_H^2/M_Z^2
\eeq
in units of $\lambda_0 = M_Z^2/v$, while the quadrilinear vertex 
carries the coefficient
\beq
\lambda_{HHHH} = 3 M_H^2/M_Z^4
\eeq
in units of $\lambda_0^2$; numerically $\lambda_0 = 33.8$~GeV. For a
Higgs mass $M_H=110$~GeV, the trilinear coupling amounts to
$\lambda_{HHH} \lambda_0/ M_Z = 1.6$ for a typical energy scale $M_Z$,
whereas the quadrilinear coupling $\lambda_{HHHH} \lambda_0^2 = 0.6$
is suppressed with respect to the trilinear coupling by a factor close
to the size of the weak gauge coupling.\s

The trilinear Higgs self-coupling can be measured directly in
pair-production of Higgs particles at hadron and high-energy $e^+ e^-$
linear colliders. Several mechanisms that are sensitive to
$\lambda_{HHH}$ can be exploited for this task. Higgs pairs can be
produced through double Higgs-strahlung off $W$ or $Z$ bosons
\cite{gounaris,ilyin}, $WW$ or $ZZ$ fusion
\citerange{ilyin,dicus}; moreover through gluon-gluon fusion
in $pp$ collisions \citerange{glover,dawson} and high-energy
$\gamma\gamma$ fusion \cite{ilyin,boudjema,nikia} at photon colliders.
The two main processes at \ee colliders are double Higgs-strahlung and
$WW$ fusion:
\beq
\begin{array}{l l l c l}
\mbox{double Higgs-strahlung}& \hspace{-0.3cm} : & e^+e^- & 
\hspace{-0.3cm} \longrightarrow &\hspace{-0.1cm}  ZHH \\
\\[-0.8cm]
& & & \hspace{-0.3cm} \scriptstyle{Z} & \\[0.1cm]
WW\ \mbox{double-Higgs fusion}& \hspace{-0.3cm} : & $\ee$ & 
\hspace{-0.3cm} \longrightarrow & \hspace{-0.1cm} \bar{\nu}_e \nu_e HH 
\\ \\[-0.8cm]
& & & \hspace{-0.3cm} \scriptstyle{WW} &
\end{array} 
\eeq
The $ZZ$ fusion process of Higgs pairs is suppressed by an order of
magnitude since the electron-$Z$ couplings are small.  Generic
diagrams for the above two processes are depicted in
Fig.~\ref{fig:diag}. \s
\begin{fmffile}{fd}
\begin{figure}
\begin{flushleft}
\underline{double Higgs-strahlung: $e^+e^-\to ZHH$}\\[1.5\baselineskip]
{\footnotesize
\unitlength1mm
\hspace{10mm}
\begin{fmfshrink}{0.7}
\begin{fmfgraph*}(24,12)
  \fmfstraight
  \fmfleftn{i}{3} \fmfrightn{o}{3}
  \fmf{fermion}{i1,v1,i3}
  \fmflabel{$e^-$}{i1} \fmflabel{$e^+$}{i3}
  \fmf{boson,lab=$Z$,lab.s=left,tens=3/2}{v1,v2}
  \fmf{boson}{v2,o3} \fmflabel{$Z$}{o3}
  \fmf{phantom}{v2,o1}
  \fmffreeze
  \fmf{dashes,lab=$H$,lab.s=right}{v2,v3} \fmf{dashes}{v3,o1}
  \fmffreeze
  \fmf{dashes}{v3,o2} 
  \fmflabel{$H$}{o2} \fmflabel{$H$}{o1}
  \fmfdot{v3}
\end{fmfgraph*}
\hspace{15mm}
\begin{fmfgraph*}(24,12)
  \fmfstraight
  \fmfleftn{i}{3} \fmfrightn{o}{3}
  \fmf{fermion}{i1,v1,i3}
  \fmf{boson,lab=$Z$,lab.s=left,tens=3/2}{v1,v2}
  \fmf{dashes}{v2,o1} \fmflabel{$H$}{o1}
  \fmf{phantom}{v2,o3}
  \fmffreeze
  \fmf{boson}{v2,v3,o3} \fmflabel{$Z$}{o3}
  \fmffreeze
  \fmf{dashes}{v3,o2} 
  \fmflabel{$H$}{o2} \fmflabel{$H$}{o1}
\end{fmfgraph*}
\hspace{15mm}
\begin{fmfgraph*}(24,12)
  \fmfstraight
  \fmfleftn{i}{3} \fmfrightn{o}{3}
  \fmf{fermion}{i1,v1,i3}
  \fmf{boson,lab=$Z$,lab.s=left,tens=3/2}{v1,v2}
  \fmf{dashes}{v2,o1} \fmflabel{$H$}{o1}
  \fmf{dashes}{v2,o2} \fmflabel{$H$}{o2}
  \fmf{boson}{v2,o3} \fmflabel{$Z$}{o3}
\end{fmfgraph*}
\end{fmfshrink}
}
\\[2\baselineskip]
\underline{$WW$ double-Higgs fusion: $e^+e^-\to \bar\nu_e\nu_e HH$}\\[1.5\baselineskip]
{\footnotesize
\unitlength1mm
\hspace{10mm}
\begin{fmfshrink}{0.7}
\begin{fmfgraph*}(24,20)
  \fmfstraight
  \fmfleftn{i}{8} \fmfrightn{o}{8}
  \fmf{fermion,tens=3/2}{i2,v1} \fmf{phantom}{v1,o2}
  \fmflabel{$e^-$}{i2}
  \fmf{phantom}{o7,v2} \fmf{fermion,tens=3/2}{v2,i7}
  \fmflabel{$e^+$}{i7}
  \fmffreeze
  \fmf{fermion}{v1,o1} \fmflabel{$\nu_e$}{o1}
  \fmf{fermion}{o8,v2} \fmflabel{$\bar\nu_e$}{o8}
  \fmf{boson}{v1,v3} 
  \fmf{boson}{v3,v2}
  \fmf{dashes,lab=$H$}{v3,v4}
  \fmf{dashes}{v4,o3} \fmf{dashes}{v4,o6}
  \fmflabel{$H$}{o3} \fmflabel{$H$}{o6}
  \fmffreeze
  \fmf{phantom,lab=$W$,lab.s=left}{v1,x1} \fmf{phantom}{x1,v3} 
  \fmf{phantom,lab=$W$,lab.s=left}{x2,v2} \fmf{phantom}{v3,x2}
  \fmfdot{v4}
\end{fmfgraph*}
\hspace{15mm}
\begin{fmfgraph*}(24,20)
  \fmfstraight
  \fmfleftn{i}{8} \fmfrightn{o}{8}
  \fmf{fermion,tens=3/2}{i2,v1} \fmf{phantom}{v1,o2}
  \fmf{phantom}{o7,v2} \fmf{fermion,tens=3/2}{v2,i7}
  \fmffreeze
  \fmf{fermion}{v1,o1}
  \fmf{fermion}{o8,v2}
  \fmf{boson}{v1,v3} 
  \fmf{boson}{v4,v2}
  \fmf{boson,lab=$W$,lab.s=left}{v3,v4}
  \fmf{dashes}{v3,o3} \fmf{dashes}{v4,o6}
  \fmflabel{$H$}{o3} \fmflabel{$H$}{o6}
\end{fmfgraph*}
\hspace{15mm}
\begin{fmfgraph*}(24,20)
  \fmfstraight
  \fmfleftn{i}{8} \fmfrightn{o}{8}
  \fmf{fermion,tens=3/2}{i2,v1} \fmf{phantom}{v1,o2}
  \fmf{phantom}{o7,v2} \fmf{fermion,tens=3/2}{v2,i7}
  \fmffreeze
  \fmf{fermion}{v1,o1}
  \fmf{fermion}{o8,v2}
  \fmf{boson}{v1,v3} 
  \fmf{boson}{v3,v2}
  \fmf{dashes}{v3,o3} \fmf{dashes}{v3,o6}
  \fmflabel{$H$}{o3} \fmflabel{$H$}{o6}
\end{fmfgraph*}
\end{fmfshrink}
}
\end{flushleft}
\caption{\textit{%
Processes contributing to Higgs-pair production in the Standard Model
at $e^+e^-$ linear colliders: double Higgs-strahlung and $WW$ fusion
(generic diagrams).
}}
\label{fig:diag}
\end{figure}
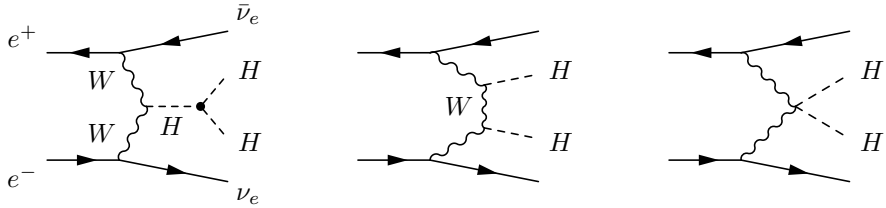\\
\indent With values typically of the order of a few fb and below, the 
cross
sections are small at \ee ~linear colliders for masses of the Higgs
boson in the intermediate mass range.  High luminosities are therefore
needed to produce a sufficiently large sample of Higgs-pair events and 
to isolate the signal from the background.\pskip

\nn
{\bf 2.} 
If light Higgs bosons with masses below about 130 GeV will be found,
the Standard Model is likely embedded in a supersymmetric theory. The
minimal supersymmetric extension of the Standard Model (MSSM) includes
two iso-doublets of Higgs fields $\varphi_1$, $\varphi_2$ which, after
three components are absorbed to provide masses to the electroweak
gauge bosons, gives rise to a quintet of physical Higgs boson states:
$h$, $H$, $A$, $H^\pm$ \cite{gunhaber}. While a strong upper bound of
about 130~GeV can be derived on the mass of the light CP-even neutral
Higgs boson $h$ \cite{okada,carena}, the heavy CP-even and CP-odd
neutral Higgs bosons $H$, $A$, and the charged Higgs bosons $H^\pm$
may have masses of the order of the electroweak symmetry scale $v$ up
to about 1~TeV. This extended Higgs system can be described by two
parameters at the tree level: one mass parameter which is generally
identified with the pseudoscalar $A$ mass $M_A$, and tan$\beta$, the
ratio of the vacuum expectation values of the two neutral fields in
the two iso-doublets.  \s

The general self-interaction potential of two Higgs doublets 
$\varphi_i$ in a CP-conserving theory
can be expressed by seven real couplings $\lambda_k$ and three real
mass parameters $m_{11}^2$, $m_{22}^2$ and $m_{12}^2$:
\beq
V[\varphi_1,\varphi_2] &=& 
m_{11}^2 \varphi_1^\dagger \varphi_1 + m_{22}^2 \varphi_2^\dagger 
\varphi_2 - [m_{12}^2 \varphi_1^\dagger \varphi_2 + {\rm h.c.}]
\non \\ 
& + & 
\textstyle{\frac{1}{2}} \lambda_1 (\varphi_1^\dagger \varphi_1)^2 + 
\textstyle{\frac{1}{2}} \lambda_2 (\varphi_2^\dagger \varphi_2)^2 +
\lambda_3 (\varphi_1^\dagger \varphi_1)(\varphi_2^\dagger \varphi_2) +
\lambda_4 (\varphi_1^\dagger \varphi_2)(\varphi_2^\dagger \varphi_1) 
\non\\
&+& \left\{ 
\textstyle{\frac{1}{2}} \lambda_5 (\varphi_1^\dagger \varphi_2)^2
+ [\lambda_6 (\varphi_1^\dagger \varphi_1) + \lambda_7
(\varphi_2^\dagger \varphi_2)]\varphi_1^\dagger \varphi_2 + 
{\rm h.c.} \right\}
\label{potential}
\eeq
In the MSSM, the $\lambda$ parameters are given at tree level by
\beq
\lambda_1 \!&=&\! \lambda_2 \;=\; \textstyle{\frac{1}{4}} (g^2 + g'^2) 
\non\\
\lambda_3 \!&=&\! \textstyle{\frac{1}{4}} (g^2 - g'^2) \non\\
\lambda_4 \!&=&\! -\textstyle{\frac{1}{2}} g^2 \non\\
\lambda_5 \!&=&\! \lambda_6 \;=\; \lambda_7 \;=\; 0
\eeq
and the mass parameters by
\beq
m_{12}^2 &=& \textstyle{\frac{1}{2}}M_A^2\sin 2\beta \non\\
m_{11}^2 &=& (M_A^2+M_Z^2) \sin^2\beta -  \textstyle{\frac{1}{2}}M_Z^2 
\non\\
m_{22}^2 &=& (M_A^2+M_Z^2) \cos^2\beta -  \textstyle{\frac{1}{2}}M_Z^2 
\eeq
The mass $M_A$ and tan$\beta$ determine the strength of the trilinear
couplings of the physical Higgs bosons, together with the electroweak
gauge couplings.

The mass parameters $m_{ij}^2$ and the couplings $\lambda_i$ in the
potential are affected by top and stop-loop radiative corrections.
Radiative corrections in the one-loop leading $m_t^4$ 
ap\-pro\-xi\-ma\-tion are parameterized by
\beq
\epsilon \approx \frac{3 G_F m_t^4}{\sqrt{2} \pi^2 \sin^2 \beta} 
\log \frac{M_S^2}{m_t^2} 
\eeq
where the scale of supersymmetry breaking is characterized by a common
squark-mass value $M_S$, set $1$~TeV in the numerical analyses; if
stop mixing effects are modest on the SUSY scale, they can be
accounted for by shifting $M_S^2$ in $\epsilon$ by the amount $\Delta
M_S^2 = \hat{A}^2 [1-\hat{A}^2/(12 M_S^2)]$. The charged and neutral
CP-even Higgs boson masses, and the mixing angle $\alpha$ are given in
this approximation by

\vspace{-0.5cm}
{\footnotesize \beq
M_{H^\pm}^2 \!\!&=&\!\!  M_A^2 + 
M_Z^2 \cos^2 \theta_W \non\\
M_{h,H}^2 \!\!&=&\!\! \textstyle{\frac{1}{2}}
\left[ M_A^2+M_Z^2+\epsilon \mp
\sqrt{(M_A^2+M_Z^2+\epsilon)^2- 4M_A^2 M_Z^2 \cos^2 2\beta
   - 4\epsilon( M_A^2 \sin^2 \beta + M_Z^2 \cos^2 \beta)} \right]
\non \\
\tan 2\alpha \!\!&=&\!\! \tan 2\beta
 \frac{M_A^2 + M_Z^2}{M_A^2 - M_Z^2 +\epsilon/\cos 2\beta} \qquad
\mbox{with} \qquad  - \frac{\pi}{2} \leq \alpha \leq 0
\label{mass}
\eeq}
\hspace{-0.3cm} when expressed in terms of the mass $M_A$ 
and tan $\beta$.\s

The set of trilinear couplings between the neutral physical Higgs 
bosons can be written \cite{okada,djouadi} in units of $\lambda_0$ as
\beq
\lambda_{hhh} &=& 3 \cos2\alpha \sin (\beta+\alpha) 
+ 3 \frac{\epsilon}{M_Z^2} \frac{\cos \alpha}{\sin\beta} \cos^2\alpha  
\non \\
\lambda_{Hhh} &=& 2\sin2 \alpha \sin (\beta+\alpha) -\cos 2\alpha \cos(\beta
+ \alpha) + 3 \frac{\epsilon}{M_Z^2} \frac{\sin \alpha}{\sin\beta}
\cos^2\alpha \non \\
\lambda_{HHh} &=& -2 \sin 2\alpha \cos (\beta+\alpha) - \cos 2\alpha \sin(\beta
+ \alpha) + 3 \frac{\epsilon}{M_Z^2} \frac{\cos \alpha}{\sin\beta}
\sin^2\alpha \non \\
\lambda_{HHH} &=& 3 \cos 2\alpha \cos (\beta+\alpha) 
+ 3 \frac{\epsilon}{M_Z^2} \frac{\sin \alpha}{\sin\beta} \sin^2 \alpha
\non \\
\lambda_{hAA} &=& \cos 2\beta \sin(\beta+ \alpha)+ 
\frac{\epsilon}{M_Z^2} 
\frac{\cos \alpha}{\sin\beta} \cos^2\beta \non \\
\lambda_{HAA} &=& - \cos 2\beta \cos(\beta+ \alpha) + 
\frac{\epsilon}{M_Z^2} 
\frac{\sin \alpha}{\sin\beta} \cos^2\beta
\label{coup}
\eeq
In the decoupling limit $M_A^2 \sim M_H^2 \sim M^2_{H^\pm} \gg v^2/2$, 
the trilinear Higgs couplings reduce to {\small \beq
\lambda_{hhh} &\lra& 3 M_h^2/M_Z^2 \non\\
\lambda_{Hhh} &\lra& -3 \sqrt{ 
\left( \frac{M_h^2}{M_Z^2}-\frac{\epsilon}{M_Z^2}\sin^2\beta \right)
\left( 1 - \frac{M_h^2}{M_Z^2} + \frac{\epsilon}{M_Z^2}\sin^2\beta 
\right) } - \frac{3\epsilon}{M_Z^2}\sin\beta\cos\beta \non\\
\lambda_{HHh} &\lra& 2 - \frac{3 M_h^2}{M_Z^2} + 
\frac{3\epsilon}{M_Z^2} \non\\
\lambda_{HHH} &\lra& 3 \sqrt{ 
\left( \frac{M_h^2}{M_Z^2}-\frac{\epsilon}{M_Z^2}\sin^2\beta \right)
\left( 1 - \frac{M_h^2}{M_Z^2} + \frac{\epsilon}{M_Z^2}\sin^2\beta 
\right) } -
\frac{3\epsilon}{M_Z^2} \frac{\cos^3\beta}{\sin\beta} \non\\
\lambda_{hAA} &\lra& - \frac{M_h^2}{M_Z^2} + \frac{\epsilon}{M_Z^2} 
\non\\
\lambda_{HAA} &\lra& \sqrt{ 
\left( \frac{M_h^2}{M_Z^2}-\frac{\epsilon}{M_Z^2}\sin^2\beta \right)
\left( 1 - \frac{M_h^2}{M_Z^2} + \frac{\epsilon}{M_Z^2}\sin^2\beta 
\right) } -
\frac{\epsilon}{M_Z^2}\frac{\cos^3\beta}{\sin\beta}
\eeq} 
\noindent \hspace{-0.3cm} with $M_h^2 = M_Z^2 \cos^2 2\beta + 
\epsilon\sin^2\beta$.  As expected, the self-coupling of the light 
CP-even neutral Higgs boson $h$ approaches the SM value in the 
decoupling limit.\s

\begin{figure}
\begin{center}
\epsfig{figure=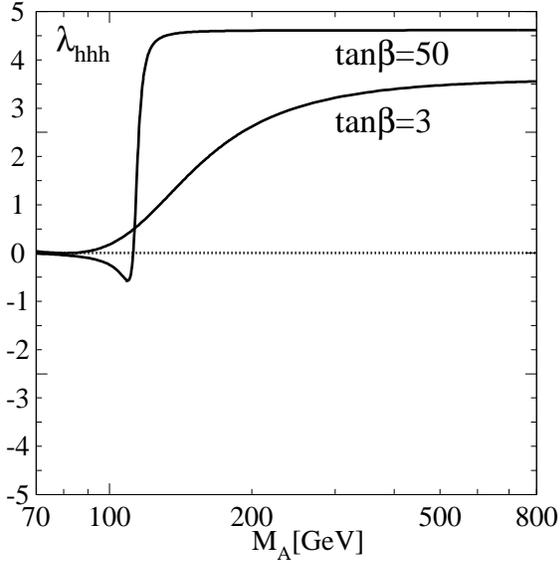,width=7.5cm}
\hspace{1cm}
\epsfig{figure=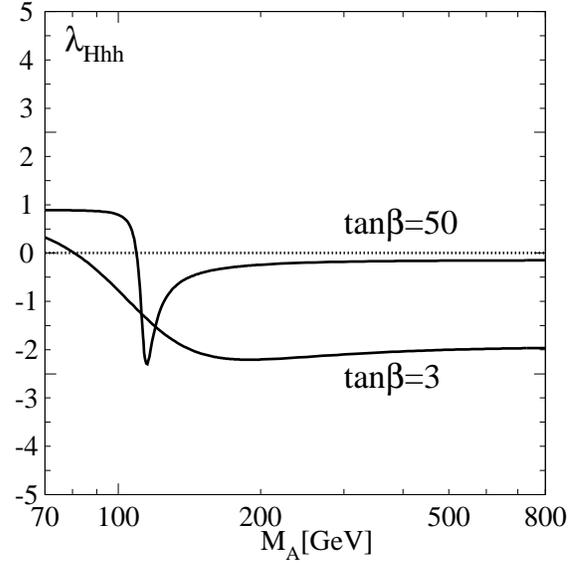,width=7.5cm}\\[2cm]
\end{center}
\begin{center}
\epsfig{figure=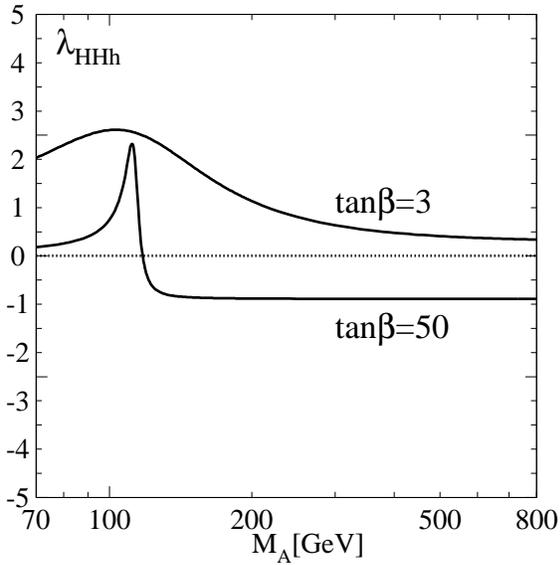,width=7.5cm}
\hspace{1cm}
\epsfig{figure=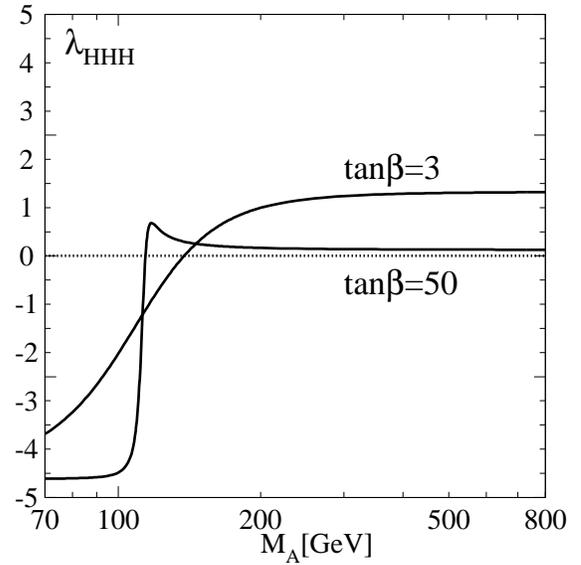,width=7.5cm}
\\[1cm]
\caption{Variation of the trilinear couplings between CP-even Higgs 
bosons with $M_A$ for tan$\beta = 3$ and $50$ in the MSSM; the region 
of rapid variations corresponds to the $h/H$ cross-over region in the 
neutral CP-even sector.}
\label{fig:lambda1}
\end{center}
\end{figure}

\begin{figure}
\begin{center}
\epsfig{figure=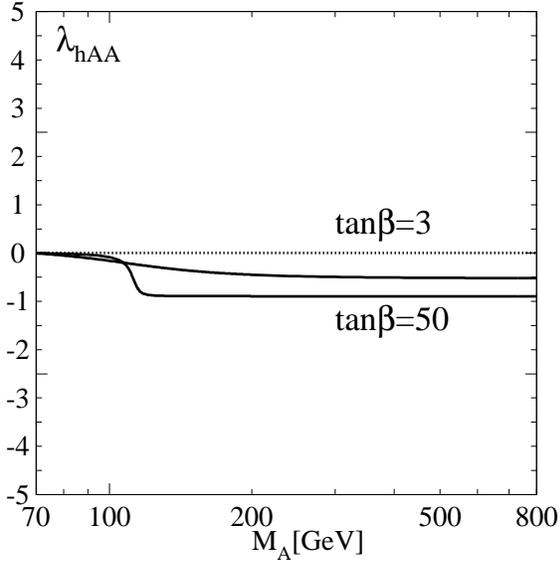,width=7.5cm}
\hspace{1cm}
\epsfig{figure=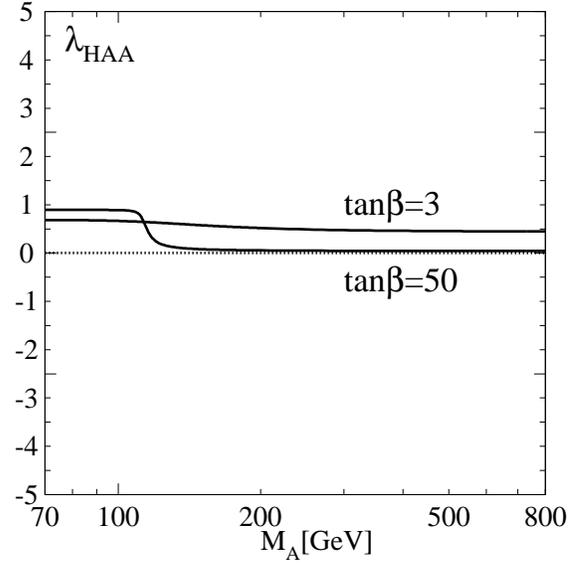,width=7.5cm}\\[2cm]
\end{center}
\begin{center}
\epsfig{figure=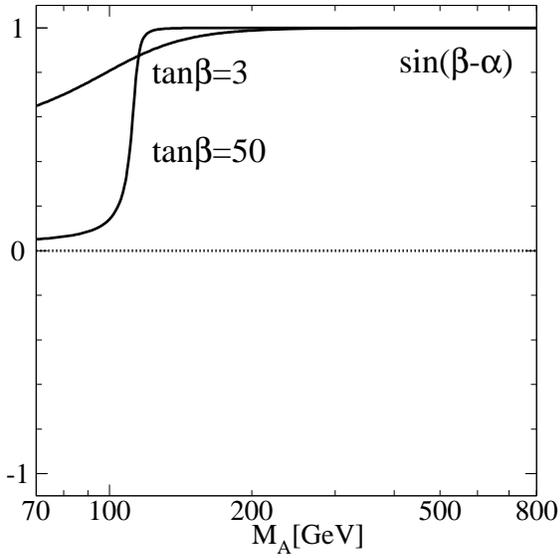,width=7.5cm}
\hspace{1cm}
\epsfig{figure=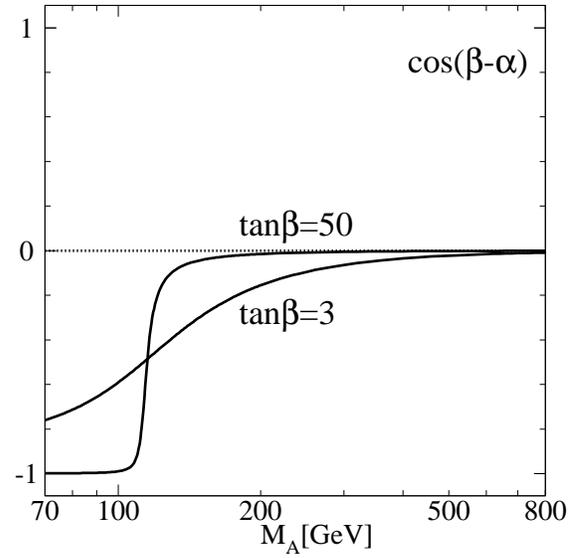,width=7.5cm}
\\[1cm]
\caption{Upper set: Variation of the trilinear scalar couplings 
between CP-even and CP-odd Higgs bosons with $M_A$ for tan$\beta = 3$ 
and $50$ in the MSSM. Lower set: ZZh and ZZH gauge couplings in units 
of the SM coupling.}
\label{fig:lambda2}
\end{center}
\end{figure}

\begin{figure}
\begin{center}
\epsfig{figure=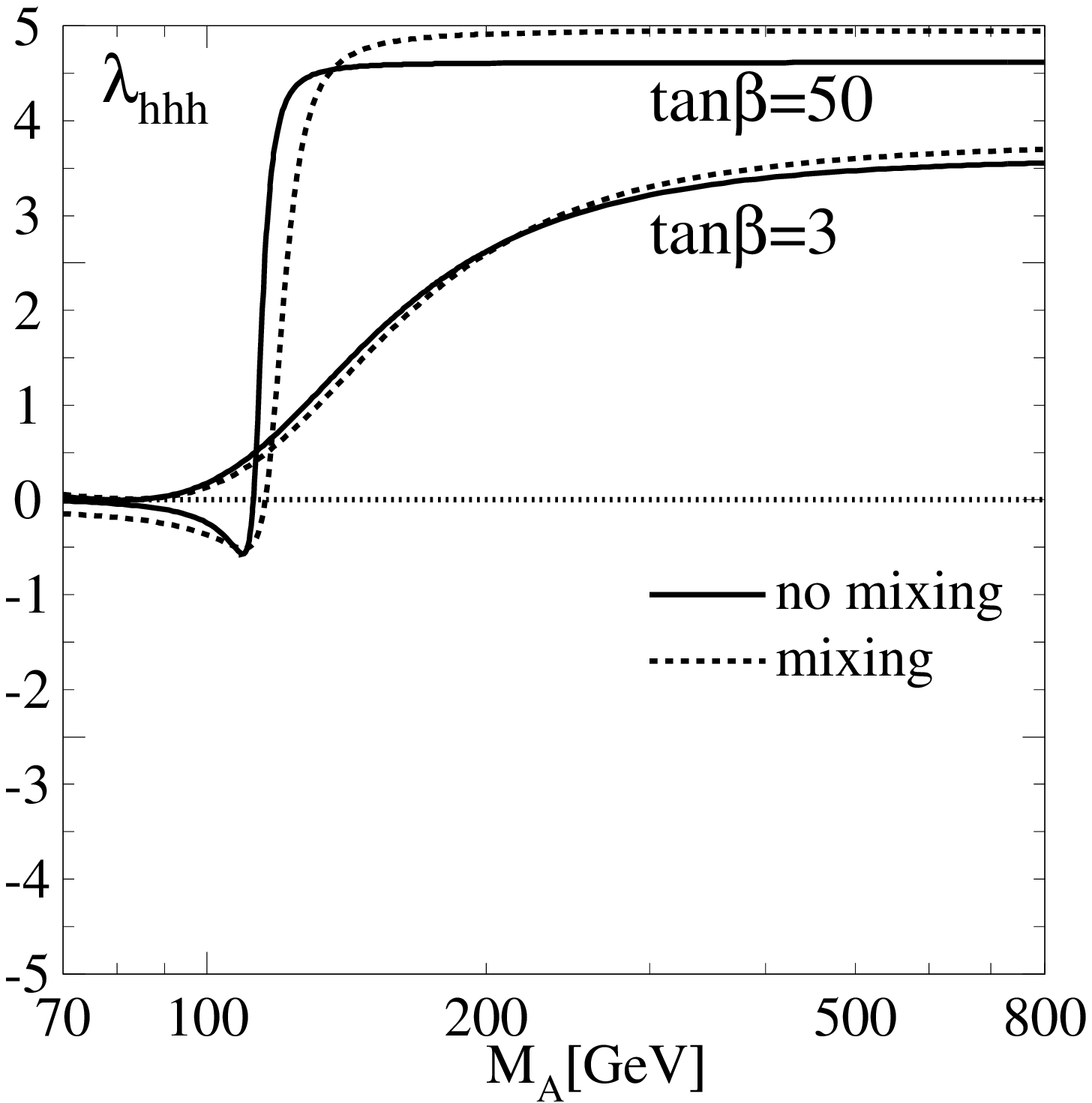,width=7.5cm}
\hspace{1cm}
\epsfig{figure=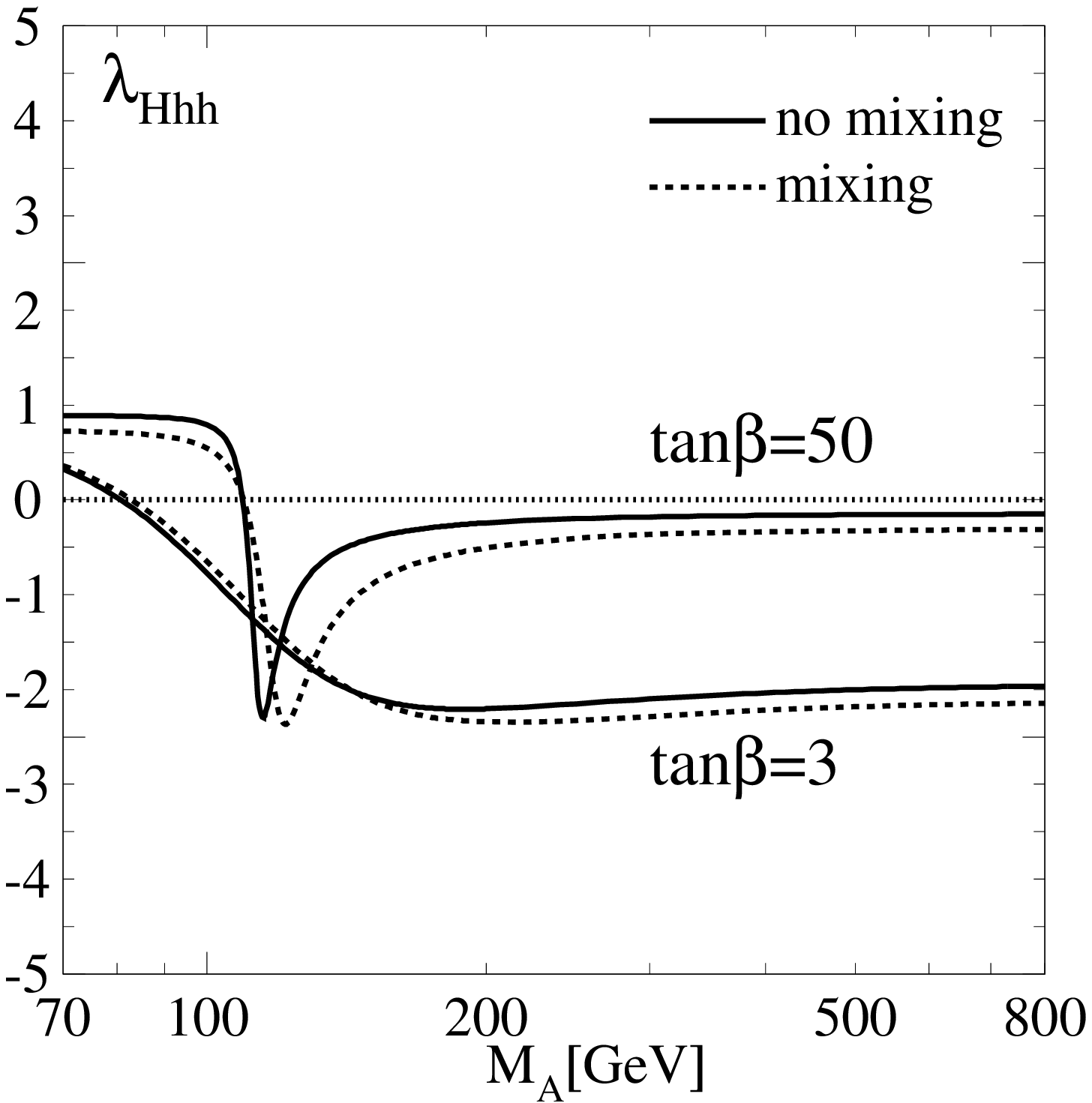,width=7.5cm}\\
\caption{Modification of the trilinear couplings $\lambda_{hhh}$ and 
$\lambda_{Hhh}$ due to mixing effects for $A=\mu=1$~TeV.}
\label{fig:coupmix}
\end{center}
\end{figure}
In the subsequent numerical analysis the complete one-loop and the
leading two-loop cor\-rec\-tions to the MSSM Higgs masses and to the
trilinear couplings are included, as presented in
Ref.~\cite{carena,hdecay}.  Mixing effects due to non-vanishing $A$,
$\mu$ parameters primarily affect the light Higgs mass; the upper
limit on $M_h$ depends strongly on the size of the mixing parameters,
raising this value for tan $\beta \gtrsim 2.5$ beyond the reach of
LEP2, cf. Ref.~\cite{carzer}. The couplings however are affected less
when evaluated for the physical Higgs masses. The variation of the
trilinear couplings with $M_A$ is shown for two values $\tan\beta = 3$
and $50$ in Figs.~\ref{fig:lambda1} and \ref{fig:lambda2}. The region
in which the couplings vary rapidly, corresponds to the $h/H$
cross-over region of the two mass branches in the neutral CP-even
sector, cf.~eq.~(\ref{mass}). The trilinear couplings between $h$, $H$
and the pseudoscalar pair $AA$ are in general significantly smaller
than the trilinear couplings among the CP-even Higgs bosons. Small
modifications of the couplings due to mixing effects are illustrated
in Fig.~\ref{fig:coupmix} (for a detailed discussion of mixing effects
see Ref.~\cite{osland}).\s

In contrast to the Standard Model, resonance production of the heavy
neutral Higgs boson $H$ followed by subsequent decays $H \to hh$,
plays a dominant role in part of the parameter space for moderate
values of $\tan\beta$ and $H$ masses between 200 and 350~GeV,
Ref.~\cite{zerwas}. In this range, the branching ratio, derived from
the partial width
\beq
\Gamma (H \to hh) = 
\frac{\sqrt{2} G_F M_Z^4 \beta_h}{32 \pi M_H} \lambda_{Hhh}^2 
\eeq
is neither too small nor too close to unity to be measured directly.
[The decay of either $h$ or $H$ into a pair of pseudoscalar states,
$h/H \to AA$, is kinematically not possible in the parameter range
which the present analysis is based upon.] If double Higgs production
is mediated by the resonant production of $H$, the total production
cross section of light Higgs pairs increases by about an order of
magnitude \cite{djouadi}.\s

The trilinear Higgs-boson couplings are involved in a large number of 
processes at \ee ~li\-ne\-ar colliders \cite{djouadi}:
\beq
\begin{array}{l@{:\quad}l@{\;\to\;}l l l l}
\mbox{double Higgs-strahlung} & $\ee$  & ZH_i H_j & \mathrm{and} 
& ZAA & [H_{i,j}=h,H] \\[0.2cm]
\mbox{triple Higgs production} & $\ee$ & AH_i H_j & \mathrm{and}
& AAA & \\[0.2cm]
WW\ \mbox{fusion} & $\ee$ & \bar{\nu}_e \nu_e H_i H_j & \mathrm{and}
& \bar{\nu}_e \nu_e AA & 
\end{array} 
\eeq
The trilinear couplings which enter for various final states, 
cf.~Fig.~\ref{fig:graphs}, are marked by a cross in the matrix 
Table~\ref{tab:coup}.
While the combination of couplings in Higgs-strahlung is isomorphic to
$WW$ fusion, it is different for associated triple Higgs production.
If all the cross sections were large enough, the system could be
solved for all $\lambda'$s, up to discrete ambiguities, based on double
Higgs-strahlung, $Ahh$ and triple $A$ production ["bottom-up
approach"]. This can easily be inferred from the correlation matrix
Table~\ref{tab:coup}. From $\sigma(ZAA)$ and $\sigma(AAA)$ the
couplings $\lambda(hAA)$ and $\lambda(HAA)$ can be extracted. In a
second step, $\sigma(Zhh)$ and $\sigma(Ahh)$ can be used to solve
for $\lambda(hhh)$ and $\lambda(Hhh)$; subsequently, $\sigma(ZHh)$ for
$\lambda(HHh)$; and, finally, $\sigma(ZHH)$ for $\lambda(HHH)$. The
remaining triple Higgs cross sections $\sigma(AHh)$ and $\sigma(AHH)$
provide additional redundant information on the trilinear couplings.\s

In practice, not all the cross sections will be large enough to be
accessible experimentally, preventing the straightforward solution for
the complete set of couplings. In this situation however the reverse
direction can be followed ["top-down approach"]. The trilinear Higgs
couplings can stringently be tested by comparing the theoretical
predictions of the cross sections with the experimental results for
the accessible channels of double and triple Higgs production. \s
\begin{table}
\begin{center}$
\renewcommand{\arraystretch}{1.3}
\begin{array}{|l||cccc|cccc|}\hline
\phantom{\lambda} & 
\multicolumn{4}{|c|}{\mathrm{double\;Higgs\!-\!strahlung}} &
\multicolumn{4}{|c|}{\phantom{d} \mathrm{triple\;Higgs\!-\!production \phantom{d}}} \str \\
\phantom{\lambda i}\lambda & Zhh & ZHh & ZHH & ZAA & \phantom{d}Ahh & \phantom{d}AHh & \phantom{d}AHH & \!\! AAA \\ \hline\hline
hhh & \times & & & & \times & & &  \\
Hhh & \times & \times & & & \times & \times & & \\
HHh & & \times & \times & & & \times & \times & \\ 
HHH & & & \times & & & & \times & \\ \hline
hAA & & & & \times & \times & \times & & \times \\ 
HAA & & & & \times & & \times & \times & \times \\
\hline
\end{array}$
\end{center}
\caption{The trilinear couplings between neutral CP-even and CP-odd 
MSSM Higgs bosons which can generically be probed in double 
Higgs-strahlung and associated triple Higgs-production, are marked by 
a cross. The matrix for WW fusion is isomorphic to the matrix for 
Higgs-strahlung.}
\label{tab:coup}
\end{table}
\begin{figure}
\begin{flushleft}
\underline{double Higgs-strahlung: $e^+e^-\to ZH_iH_j$, $ZAA$
[$H_{i,j}=h,H$]}\\[1.5\baselineskip]
{\footnotesize
\unitlength1mm
\hspace{5mm}
\begin{fmfshrink}{0.7}
\begin{fmfgraph*}(24,12)
  \fmfstraight
  \fmfleftn{i}{3} \fmfrightn{o}{3}
  \fmf{fermion}{i1,v1,i3}
  \fmf{boson,tens=3/2}{v1,v2}
  \fmf{boson}{v2,o3} \fmflabel{$Z$}{o3}
  \fmf{phantom}{v2,o1}
  \fmffreeze
  \fmf{dashes,lab=$H_{i,,j}$,lab.s=right}{v2,v3} \fmf{dashes}{v3,o1}
  \fmffreeze
  \fmf{dashes}{v3,o2} 
  \fmflabel{$H_{i,j}$}{o2} \fmflabel{$H_{i,j}$}{o1}
  \fmfdot{v3}
\end{fmfgraph*}
\hspace{15mm}
\begin{fmfgraph*}(24,12)
  \fmfstraight
  \fmfleftn{i}{3} \fmfrightn{o}{3}
  \fmf{fermion}{i1,v1,i3}
  \fmf{boson,tens=3/2}{v1,v2}
  \fmf{dashes}{v2,o1} \fmflabel{$H$}{o1}
  \fmf{phantom}{v2,o3}
  \fmffreeze
  \fmf{dashes,lab=$A$,lab.s=left}{v2,v3} 
  \fmf{boson}{v3,o3} \fmflabel{$Z$}{o3}
  \fmffreeze
  \fmf{dashes}{v3,o2} 
  \fmflabel{$H_{i,j}$}{o2} \fmflabel{$H_{i,j}$}{o1}
\end{fmfgraph*}
\hspace{15mm}
\begin{fmfgraph*}(24,12)
  \fmfstraight
  \fmfleftn{i}{3} \fmfrightn{o}{3}
  \fmf{fermion}{i1,v1,i3}
  \fmf{boson,tens=3/2}{v1,v2}
  \fmf{dashes}{v2,o1} \fmflabel{$H$}{o1}
  \fmf{phantom}{v2,o3}
  \fmffreeze
  \fmf{boson}{v2,v3,o3} \fmflabel{$Z$}{o3}
  \fmffreeze
  \fmf{dashes}{v3,o2} 
  \fmflabel{$H_{i,j}$}{o2} \fmflabel{$H_{i,j}$}{o1}
\end{fmfgraph*}
\hspace{15mm}
\begin{fmfgraph*}(24,12)
  \fmfstraight
  \fmfleftn{i}{3} \fmfrightn{o}{3}
  \fmf{fermion}{i1,v1,i3}
  \fmf{boson,tens=3/2}{v1,v2}
  \fmf{dashes}{v2,o1} \fmflabel{$H_{i,j}$}{o1}
  \fmf{dashes}{v2,o2} \fmflabel{$H_{i,j}$}{o2}
  \fmf{boson}{v2,o3} \fmflabel{$Z$}{o3}
\end{fmfgraph*}
\\[2\baselineskip]
\hspace{5mm}
\begin{fmfgraph*}(24,12)
  \fmfstraight
  \fmfleftn{i}{3} \fmfrightn{o}{3}
  \fmf{fermion}{i1,v1,i3}
  \fmf{boson,tens=3/2}{v1,v2}
  \fmf{boson}{v2,o3} \fmflabel{$Z$}{o3}
  \fmf{phantom}{v2,o1}
  \fmffreeze
  \fmf{dashes,lab=$H_{i,,j}$,lab.s=right}{v2,v3} \fmf{dashes}{v3,o1}
  \fmffreeze
  \fmf{dashes}{v3,o2} 
  \fmflabel{$A$}{o2} \fmflabel{$A$}{o1}
  \fmfdot{v3}
\end{fmfgraph*}
\hspace{15mm}
\begin{fmfgraph*}(24,12)
  \fmfstraight
  \fmfleftn{i}{3} \fmfrightn{o}{3}
  \fmf{fermion}{i1,v1,i3}
  \fmf{boson,tens=3/2}{v1,v2}
  \fmf{dashes}{v2,o1} \fmflabel{$H$}{o1}
  \fmf{phantom}{v2,o3}
  \fmffreeze
  \fmf{dashes,lab=$H_{i,,j}$,lab.s=left}{v2,v3} 
  \fmf{boson}{v3,o3} \fmflabel{$Z$}{o3}
  \fmffreeze
  \fmf{dashes}{v3,o2} 
  \fmflabel{$A$}{o2} \fmflabel{$A$}{o1}
\end{fmfgraph*}
\hspace{15mm}
\begin{fmfgraph*}(24,12)
  \fmfstraight
  \fmfleftn{i}{3} \fmfrightn{o}{3}
  \fmf{fermion}{i1,v1,i3}
  \fmf{boson,tens=3/2}{v1,v2}
  \fmf{dashes}{v2,o1} \fmflabel{$A$}{o1}
  \fmf{dashes}{v2,o2} \fmflabel{$A$}{o2}
  \fmf{boson}{v2,o3} \fmflabel{$Z$}{o3}
\end{fmfgraph*}
\end{fmfshrink}
}
\\[2\baselineskip]
\underline{triple Higgs production: $e^+e^-\to AH_iH_j$, $AAA$}
\\[1.5\baselineskip]
{\footnotesize
\unitlength1mm
\hspace{5mm}
\begin{fmfshrink}{0.7}
\begin{fmfgraph*}(24,12)
  \fmfstraight
  \fmfleftn{i}{3} \fmfrightn{o}{3}
  \fmf{fermion}{i1,v1,i3}
  \fmf{boson,tens=3/2}{v1,v2}
  \fmf{dashes}{v2,o3} \fmflabel{$A$}{o3}
  \fmf{phantom}{v2,o1}
  \fmffreeze
  \fmf{dashes,lab=$H_{i,,j}$,lab.s=right}{v2,v3} \fmf{dashes}{v3,o1}
  \fmffreeze
  \fmf{dashes}{v3,o2} 
  \fmflabel{$H_{i,j}$}{o2} \fmflabel{$H_{i,j}$}{o1}
  \fmfdot{v3}
\end{fmfgraph*}
\hspace{15mm}
\begin{fmfgraph*}(24,12)
  \fmfstraight
  \fmfleftn{i}{3} \fmfrightn{o}{3}
  \fmf{fermion}{i1,v1,i3}
  \fmf{boson,tens=3/2}{v1,v2}
  \fmf{dashes}{v2,o1} \fmflabel{$H_{i,j}$}{o1}
  \fmf{phantom}{v2,o3}
  \fmffreeze
  \fmf{dashes,lab=$A$,lab.s=left}{v2,v3} 
  \fmf{dashes}{v3,o3} \fmflabel{$A$}{o3}
  \fmffreeze
  \fmf{dashes}{v3,o2} 
  \fmflabel{$H_{i,j}$}{o2} \fmflabel{$A$}{o3}
\end{fmfgraph*}
\hspace{15mm}
\begin{fmfgraph*}(24,12)
  \fmfstraight
  \fmfleftn{i}{3} \fmfrightn{o}{3}
  \fmf{fermion}{i1,v1,i3}
  \fmf{boson,tens=3/2}{v1,v2}
  \fmf{dashes}{v2,o1}
  \fmf{phantom}{v2,o3}
  \fmffreeze
  \fmf{boson}{v2,v3} 
  \fmf{dashes}{v3,o3} \fmflabel{$A$}{o3}
  \fmffreeze
  \fmf{dashes}{v3,o2} 
  \fmflabel{$H_{i,j}$}{o2} \fmflabel{$H_{i,j}$}{o1}
\end{fmfgraph*}
\\[2\baselineskip]
\hspace{5mm}
\begin{fmfgraph*}(24,12)
  \fmfstraight
  \fmfleftn{i}{3} \fmfrightn{o}{3}
  \fmf{fermion}{i1,v1,i3}
  \fmf{boson,tens=3/2}{v1,v2}
  \fmf{dashes}{v2,o3} \fmflabel{$A$}{o3}
  \fmf{phantom}{v2,o1}
  \fmffreeze
  \fmf{dashes,lab=$H_{i,,j}$,lab.s=right}{v2,v3} \fmf{dashes}{v3,o1}
  \fmffreeze
  \fmf{dashes}{v3,o2} 
  \fmflabel{$A$}{o2} \fmflabel{$A$}{o1}
  \fmfdot{v3}
\end{fmfgraph*}
\end{fmfshrink}
}
\\[2\baselineskip]
\underline{$WW$ fusion: $e^+e^-\to\bar\nu_e\nu_e H_i H_j$, $AA$}
\\[1.5\baselineskip]
{\footnotesize
\unitlength1mm
\hspace{5mm}
\begin{fmfshrink}{0.7}
\begin{fmfgraph*}(24,20)
  \fmfstraight
  \fmfleftn{i}{8} \fmfrightn{o}{8}
  \fmf{fermion,tens=3/2}{i2,v1} \fmf{phantom}{v1,o2}
  \fmf{phantom}{o7,v2} \fmf{fermion,tens=3/2}{v2,i7}
  \fmffreeze
  \fmf{fermion}{v1,o1}
  \fmf{fermion}{o8,v2}
  \fmf{boson}{v1,v3} 
  \fmf{boson}{v3,v2}
  \fmf{dashes,lab=$H_{i,,j}$}{v3,v4}
  \fmf{dashes}{v4,o3} \fmf{dashes}{v4,o6}
  \fmflabel{$H_{i,j}$}{o3} \fmflabel{$H_{i,j}$}{o6}
  \fmffreeze
  \fmf{phantom,lab=$W$,lab.s=left}{v1,x1} \fmf{phantom}{x1,v3} 
  \fmf{phantom,lab=$W$,lab.s=left}{x2,v2} \fmf{phantom}{v3,x2}
  \fmfdot{v4}
\end{fmfgraph*}
\hspace{15mm}
\begin{fmfgraph*}(24,20)
  \fmfstraight
  \fmfleftn{i}{8} \fmfrightn{o}{8}
  \fmf{fermion,tens=3/2}{i2,v1} \fmf{phantom}{v1,o2}
  \fmf{phantom}{o7,v2} \fmf{fermion,tens=3/2}{v2,i7}
  \fmffreeze
  \fmf{fermion}{v1,o1}
  \fmf{fermion}{o8,v2}
  \fmf{boson}{v1,v3} 
  \fmf{boson}{v4,v2}
  \fmf{boson,lab=$W$,lab.s=left}{v3,v4}
  \fmf{dashes}{v3,o3} \fmf{dashes}{v4,o6}
  \fmflabel{$H_{i,j}$}{o3} \fmflabel{$H_{i,j}$}{o6}
\end{fmfgraph*}
\hspace{15mm}
\begin{fmfgraph*}(24,20)
  \fmfstraight
  \fmfleftn{i}{8} \fmfrightn{o}{8}
  \fmf{fermion,tens=3/2}{i2,v1} \fmf{phantom}{v1,o2}
  \fmf{phantom}{o7,v2} \fmf{fermion,tens=3/2}{v2,i7}
  \fmffreeze
  \fmf{fermion}{v1,o1}
  \fmf{fermion}{o8,v2}
  \fmf{boson}{v1,v3} 
  \fmf{boson}{v4,v2}
  \fmf{dashes,lab=$H^\pm$,lab.s=left}{v3,v4}
  \fmf{dashes}{v3,o3} \fmf{dashes}{v4,o6}
  \fmflabel{$H_{i,j}$}{o3} \fmflabel{$H_{i,j}$}{o6}
\end{fmfgraph*}
\hspace{15mm}
\begin{fmfgraph*}(24,20)
  \fmfstraight
  \fmfleftn{i}{8} \fmfrightn{o}{8}
  \fmf{fermion,tens=3/2}{i2,v1} \fmf{phantom}{v1,o2}
  \fmf{phantom}{o7,v2} \fmf{fermion,tens=3/2}{v2,i7}
  \fmffreeze
  \fmf{fermion}{v1,o1}
  \fmf{fermion}{o8,v2}
  \fmf{boson}{v1,v3} 
  \fmf{boson}{v3,v2}
  \fmf{dashes}{v3,o3} \fmf{dashes}{v3,o6}
  \fmflabel{$H_{i,j}$}{o3} \fmflabel{$H_{i,j}$}{o6}
\end{fmfgraph*}
\\[2\baselineskip]
\hspace{5mm}
\begin{fmfgraph*}(24,20)
  \fmfstraight
  \fmfleftn{i}{8} \fmfrightn{o}{8}
  \fmf{fermion,tens=3/2}{i2,v1} \fmf{phantom}{v1,o2}
  \fmf{phantom}{o7,v2} \fmf{fermion,tens=3/2}{v2,i7}
  \fmffreeze
  \fmf{fermion}{v1,o1}
  \fmf{fermion}{o8,v2}
  \fmf{boson}{v1,v3} 
  \fmf{boson}{v3,v2}
  \fmf{dashes,lab=$H_{i,,j}$}{v3,v4}
  \fmf{dashes}{v4,o3} \fmf{dashes}{v4,o6}
  \fmflabel{$A$}{o3} \fmflabel{$A$}{o6}
  \fmfdot{v4}
\end{fmfgraph*}
\hspace{15mm}
\begin{fmfgraph*}(24,20)
  \fmfstraight
  \fmfleftn{i}{8} \fmfrightn{o}{8}
  \fmf{fermion,tens=3/2}{i2,v1} \fmf{phantom}{v1,o2}
  \fmf{phantom}{o7,v2} \fmf{fermion,tens=3/2}{v2,i7}
  \fmffreeze
  \fmf{fermion}{v1,o1}
  \fmf{fermion}{o8,v2}
  \fmf{boson}{v1,v3} 
  \fmf{boson}{v4,v2}
  \fmf{dashes,lab=$H^\pm$,lab.s=left}{v3,v4}
  \fmf{dashes}{v3,o3} \fmf{dashes}{v4,o6}
  \fmflabel{$A$}{o3} \fmflabel{$A$}{o6}
\end{fmfgraph*}
\hspace{15mm}
\begin{fmfgraph*}(24,20)
  \fmfstraight
  \fmfleftn{i}{8} \fmfrightn{o}{8}
  \fmf{fermion,tens=3/2}{i2,v1} \fmf{phantom}{v1,o2}
  \fmf{phantom}{o7,v2} \fmf{fermion,tens=3/2}{v2,i7}
  \fmffreeze
  \fmf{fermion}{v1,o1}
  \fmf{fermion}{o8,v2}
  \fmf{boson}{v1,v3} 
  \fmf{boson}{v3,v2}
  \fmf{dashes}{v3,o3} \fmf{dashes}{v3,o6}
  \fmflabel{$A$}{o3} \fmflabel{$A$}{o6}
\end{fmfgraph*}
\end{fmfshrink}
}
\end{flushleft}
\caption{\textit{%
Processes contributing to double and triple Higgs production involving
trilinear couplings in the MSSM.
}}
\label{fig:graphs}
\end{figure}
\end{fmffile}

The processes \ee$\to ZH_i A$ and \ee$\to\bar{\nu}_e \nu_eH_i A$
[$H_i=h,$ $H$] of mixed CP-even/CP-odd Higgs final states are
generated through gauge interactions alone, mediated by virtual $Z$
bosons decaying to the CP even--odd Higgs pair, $Z^* \to H_i A$. These
parity-mixed processes do not involve trilinear Higgs-boson couplings.
\pskip

\nn
{\bf 3.}
In this paper we compare different mechanisms for the production of
Higgs boson pairs in the Standard Model and in the minimal
supersymmetric extension. An excerpt of the results, including
comparisons with LHC channels, has been published in
Ref.~\cite{muehl}. The relation to general 2-Higgs doublet models has
recently been discussed in Ref.~\cite{dubinin}. The analyses have been
carried out for $e^+ e^-$ linear colliders \cite{acco}, which are
currently designed \cite{huebner} for a low-energy phase in the range
$\sqrt{s} = 500$~GeV to 1~TeV, and a high-energy phase above 1~TeV,
potentially extending up to 5~TeV. The small cross sections require
high luminosities as foreseen in the TESLA design with targets of
$\int {\cal L} = 300$ and 500~fb$^{-1}$ {\it per annum} for $\sqrt{s}
= 500$ and $800$~GeV, respectively \cite{brink}. By analyzing the
entire ensemble of multi-Higgs final states as defined
in~Ref.\cite{djouadi}, a theoretically coherent picture has been
developed for testing the trilinear self-couplings at high-energy
colliders. Moreover, the fusion processes are calculated exactly
without reference to the equivalent-particle approximation.
Experimental simulations of signal and background processes depend
strongly on detector properties; they are beyond the scope of the
present study which describes the first modest theoretical steps into
this area. Crude estimates for final states \ee $\to Z (b\bar{b})
(b\bar{b})$ \cite{moretti} and \ee $\to Z (WW) (WW)$ \cite{kalin} can
however be derived from the existing literature; dedicated analyses of
the reducible $Z (b\bar{b}) (b\bar{b})$ background channel will be
available in the near future \cite{millermor}.\s

The paper is divided into two parts. In Section 2 we discuss the
measurement of the trilinear Higgs coupling in the Standard Model for
double Higgs-strahlung and $WW$ fusion at \ee ~linear colliders. In
Section 3 this program, including the triple Higgs production, is
extended to the Minimal Supersymmetric Standard Model MSSM.

\subsection*{2. Higgs Pair--Production in the Standard Model}

\subsubsection*{2.1 Double Higgs-strahlung}

The (unpolarized) differential cross section for the process of double
Higgs-strahlung \ee $\to ZHH$, cf.~Fig.~\ref{fig:diag}, can be cast
into the form \cite{djouadi}
\beq 
\frac{d \sigma (e^+ e^- \to ZHH)}{d x_1 d x_2} = 
\frac{\sqrt{2} G_F^3 M_Z^6}{384 \pi^3 s}
\frac{v_e^2 + a_e^2}{(1- \mu_Z)^2}\, {\cal Z} 
\eeq 
after the angular dependence is integrated out. The vector and
axial-vector $Z$ charges of the electron are defined as usual, by $v_e
= -1 + 4\sin^2 \theta_W$ and $a_e = -1$. $x_{1,2} = 2
E_{1,2}/\sqrt{s}$ are the scaled energies of the two Higgs particles,
$x_3 = 2 - x_1 -x_2$ is the scaled energy of the $Z$ boson, and $y_i =
1 - x_i$; the square of the reduced masses is denoted by $\mu_i =
M_i^2/s$, and $\mu_{ij}=\mu_i-\mu_j$. In terms of these variables,
the coefficient ${\cal Z}$ may be written as:
\beq 
{\cal Z} =  {\mathfrak a}^2 f_0 +
\frac{1}{4 \mu_Z (y_1+\mu_{HZ})} \left[ 
\frac{f_1}{y_1+\mu_{HZ}} + \frac{f_2}{y_2+\mu_{HZ}} 
+ 2\mu_Z {\mathfrak a} f_3 \right] 
+ \Bigg\{ y_1 \leftrightarrow y_2 \Bigg\} 
\eeq
with
\beq
{\mathfrak a} = \frac{\lambda_{HHH}}{y_3-\mu_{HZ}}
 + \frac{2}{y_1+\mu_{HZ}} + 
\frac{2}{y_2+\mu_{HZ}} + \frac{1}{\mu_Z}
\eeq
The coefficients $f_i$ are given by the following expressions,
\beq
f_0 &=& \mu_Z[(y_1+y_2)^2 + 8\mu_Z]/8 \non\\
f_1 &=& (y_1-1)^2(\mu_Z-y_1)^2-4\mu_H y_1(y_1+y_1\mu_Z-4\mu_Z) \non\\
& & {} + \mu_Z(\mu_Z-4\mu_H)(1-4\mu_H)-\mu_Z^2  
\non\\
f_2 &=& [\mu_Z(y_3+\mu_Z - 8\mu_H)-(1+\mu_Z)y_1 y_2](1+y_3+2\mu_Z) 
\non\\
& & {}+ y_1 y_2[y_1 y_2 + 1 + \mu_Z^2+4\mu_H (1+\mu_Z)]
+ 4\mu_H \mu_Z(1+\mu_Z+4\mu_H)+ \mu_Z^2 
\non\\
f_3 &=& y_1(y_1-1)(\mu_Z-y_1)-y_2(y_1+1)(y_1+\mu_Z)+2\mu_Z
(1+\mu_Z-4\mu_H) 
\eeq
The first term in the coefficient ${\mathfrak a}$ includes the trilinear
coupling $\lambda_{HHH}$. The other terms are related to sequential
Higgs-strahlung amplitudes and the 4-gauge-Higgs boson coupling; the 
individual terms can easily be identified by examining the 
characteristic propagators. \s
\begin{figure}
\begin{center}
\epsfig{figure=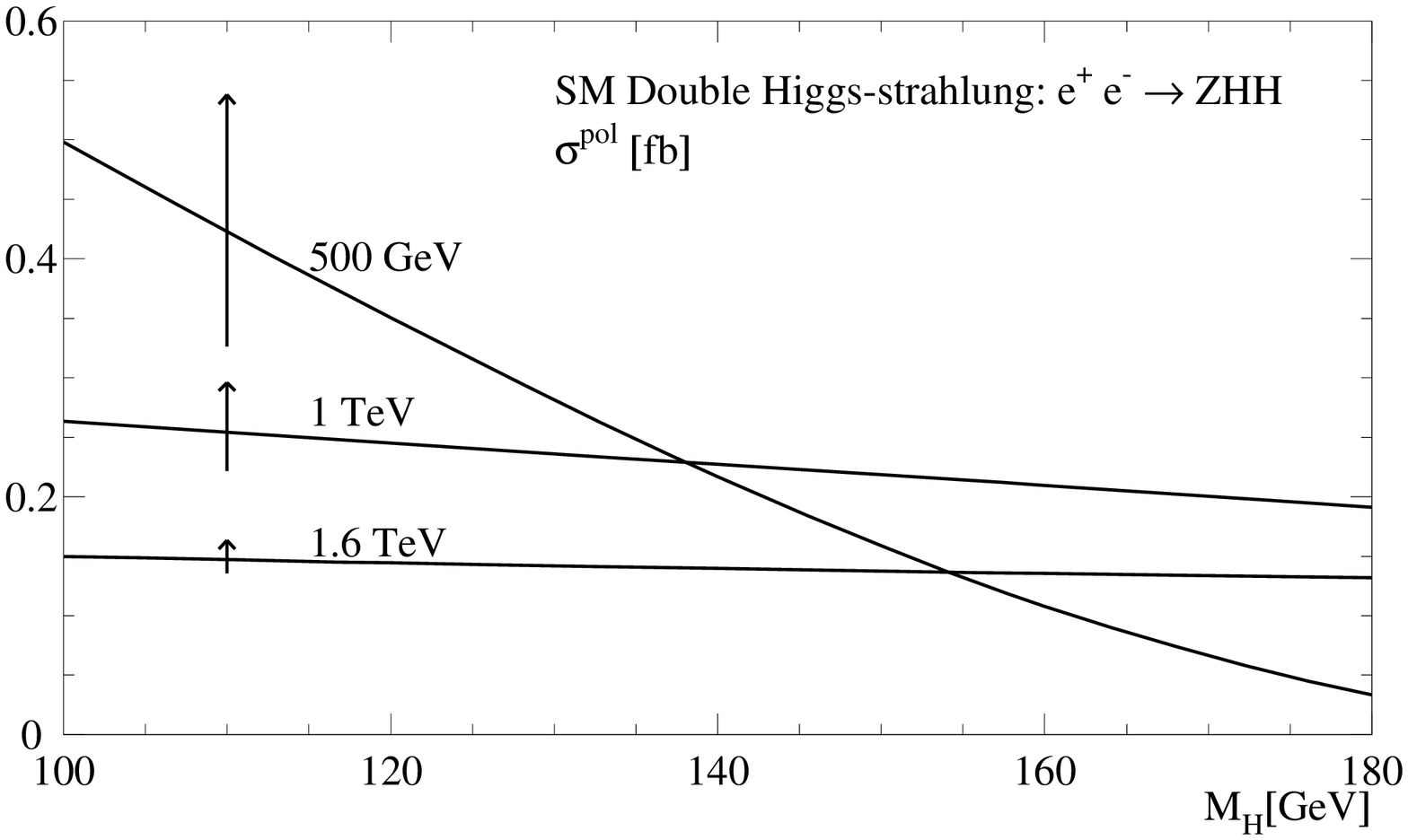,width=13cm}
\\
\end{center}
Figure 6a: {\it The cross section for double Higgs-strahlung in the SM 
at three collider energies: $500$~GeV, $1$~TeV and $1.6$~TeV. The 
electron/positron beams are taken oppositely polarized. The vertical 
arrows correspond to a variation of the trilinear Higgs coupling from 
$1/2$ to $3/2$ of the SM value.}
\label{fig:SM1}
\end{figure}
\begin{figure}
\begin{center}
\epsfig{figure=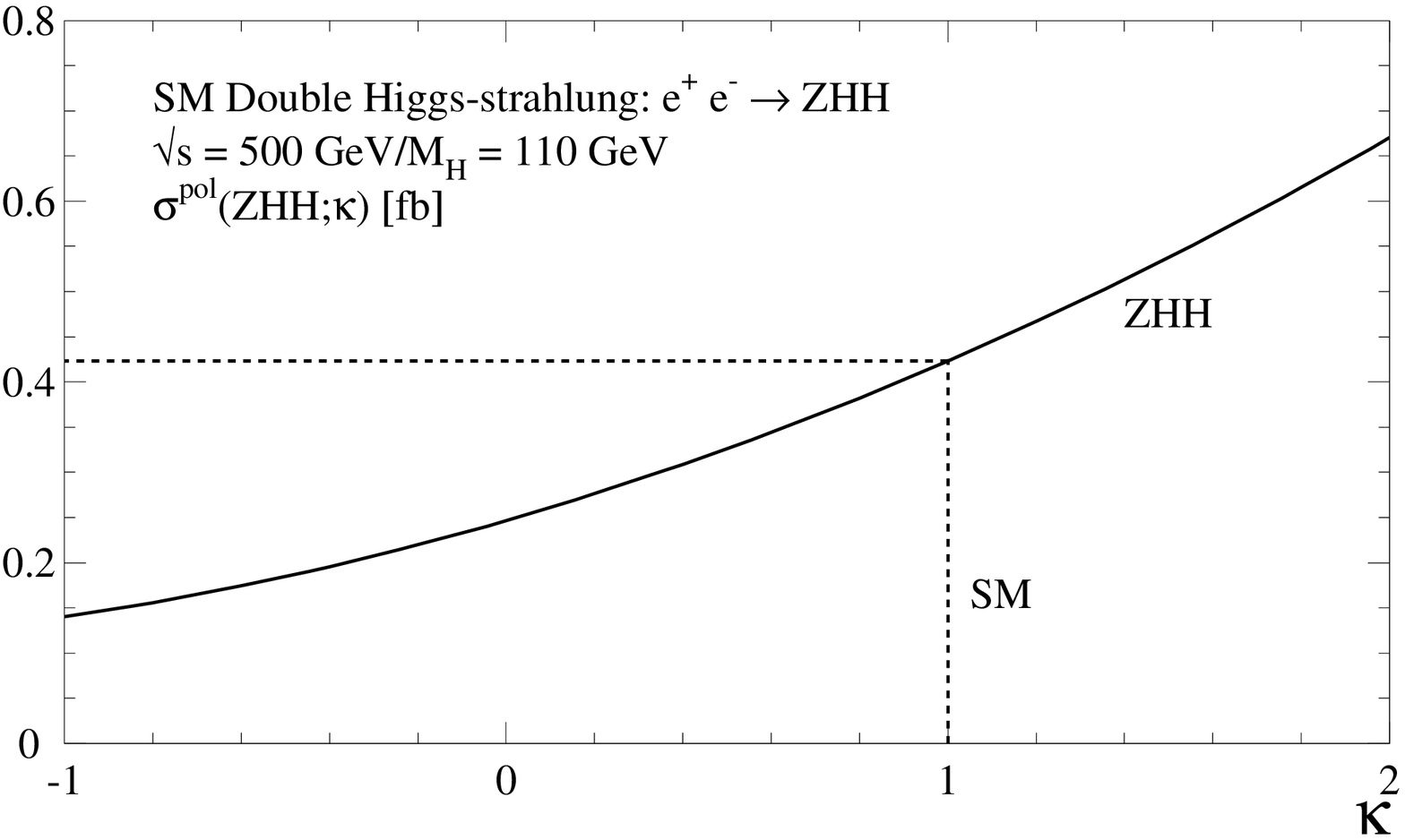,width=13cm}
\end{center}
Figure 6b: {\it Variation of the cross section $\sigma (ZHH)$ with the 
modified trilinear coupling $\kappa \lambda_{HHH}$ at a collider 
energy of $\sqrt{s}=500$~GeV and $M_H=110$~GeV.}
\label{fig:smvar1}
\end{figure}
\begin{figure}
\begin{center}
\epsfig{figure=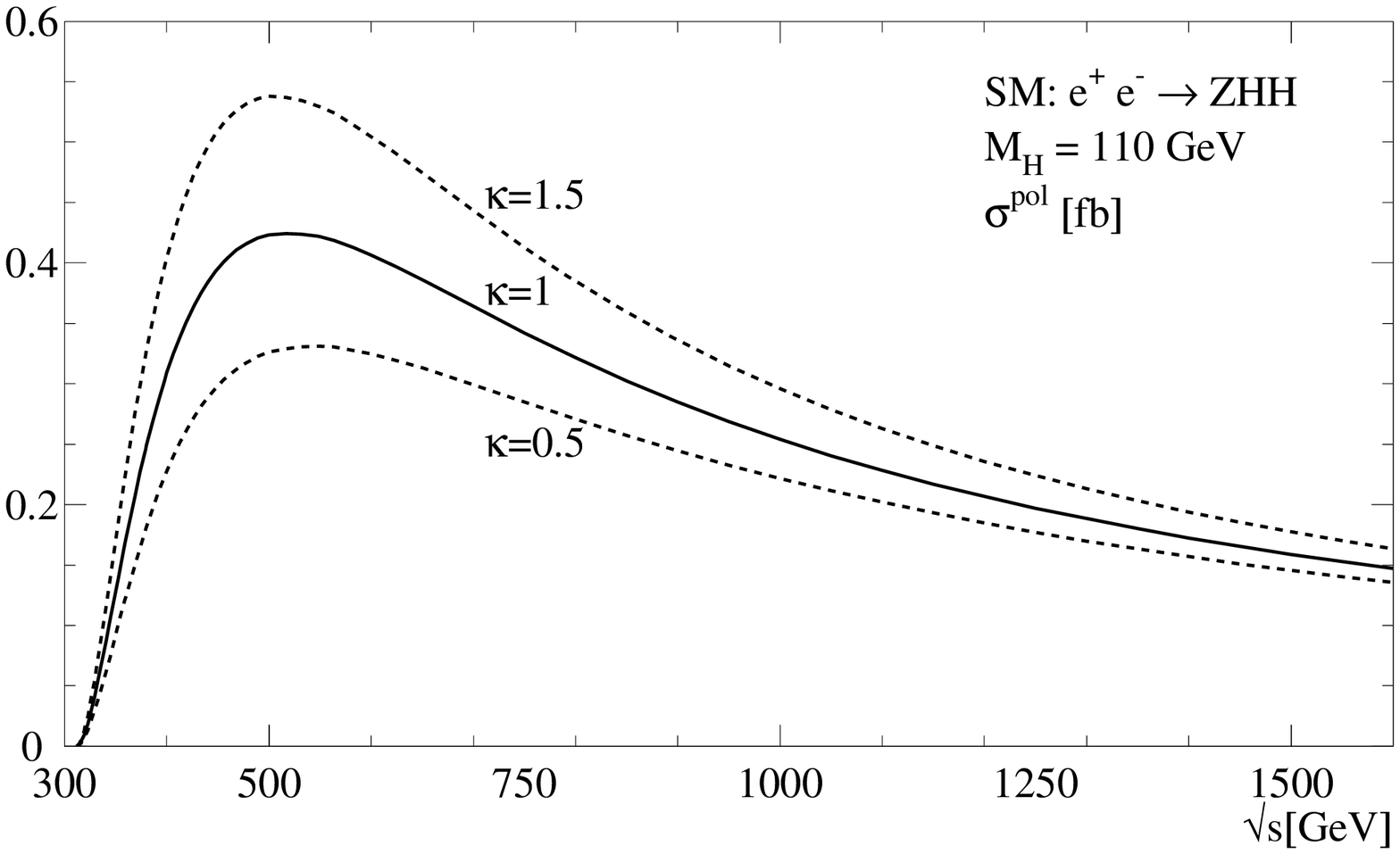,width=13cm}
\end{center}
Figure 6c: {\it The energy dependence of the cross section for double 
Higgs-strahlung for a fixed Higgs mass $M_H=110$~GeV. The variation of 
the cross section for modified trilinear couplings 
$\kappa\lambda_{HHH}$ is indicated by the dashed lines.}
\end{figure}

Since double Higgs-strahlung is mediated by s-channel $Z$-boson
exchange, the cross section doubles if oppositely polarized electron
and positron beams are used. \s

The cross sections for double Higgs-strahlung in the intermediate mass
range are presented in Fig.~6a for total $e^+ e^-$ energies of
$\sqrt{s} = 500$~GeV, 1~TeV and 1.6~TeV. The cross sections are shown
for polarized electrons and positrons [$\lambda_{e^-} \lambda_{e^+} =
-1$]; they reduce by a factor of 2 for unpolarized beams. As a result
of the scaling behavior, the cross section for double Higgs-strahlung
decreases with rising energy beyond the threshold region. The cross
section increases with rising trilinear self-coupling in the vicinity
of the SM value. The sensitivity to the $HHH$ self-coupling is
demonstrated in Fig.~6b for $\sqrt{s}=$ 500~GeV and
$M_H=$ 110~GeV by varying the trilinear coupling $\kappa\lambda_{HHH}$
within the range $\kappa = -1$ and $+2$; the sensitivity is also
illustrated by the vertical arrows in Fig.~6a for a variation $\kappa$
between 1/2 and 3/2. Evidently the cross section $\sigma($\ee$\to
ZHH)$ is sensitive to the value of the trilinear coupling, which is not 
swamped by the irreducible background diagrams involving only the
Higgs-gauge couplings. While the irreducible background diagrams
become more important for rising energies, the sensitivity to the
trilinear Higgs coupling is very large just above the kinematical
threshold for the $ZHH$ final state as demonstrated in Fig.~6c. Near
the threshold the propagator of the intermediate virtual Higgs boson
connecting to the two real Higgs bosons through $\lambda_{HHH}$ in the
final state is maximal. The maximum cross section for double
Higgs-strahlung is reached at energies $\sqrt{s}\sim
2M_H+M_Z+200$~GeV, i.e. for Higgs masses in the lower part of the
intermediate range at $\sqrt{s} \sim 500$~GeV.\s

\subsubsection*{2.2 $WW$ Double-Higgs Fusion}

The $WW$ fusion mechanism in $e^+ e^- \to \bar{\nu}_e \nu_e HH$,
cf.~Fig.~\ref{fig:diag}, provides the largest cross section for Higgs
bosons pairs in the intermediate mass range at high $e^+ e^-$~collider 
energies, in particular for polarized beams.\s

The fusion cross section can roughly be estimated in the
equivalent $W$-boson ap\-pro\-xi\-ma\-tion. The production amplitude 
for the dominant longitudinal degrees of freedom is given 
\cite{kallian} by
\beq
\hspace{-0.8cm}
{\cal M}_{LL} = \,\scriptstyle{\frac{G_F \hat{s}}{\sqrt{2}}} 
\, \left\{ \textstyle{(1 \!+ \beta_W^2)} \left[ 1 \! + \, 
\scriptstyle{\frac{\lambda_{HHH}}{(\hat{s}-M_H^2)/M_Z^2}} \right] 
\! + \,\scriptstyle{\frac{1}{\beta_W \beta_H}} \left[ 
\scriptstyle{\frac{(1-\beta_W^4)+ 
(\beta_W - \beta_H \cos\theta)^2}{\cos\theta - x_W}} \,\textstyle{-}\,
\scriptstyle{\frac{(1-\beta_W^4) + 
(\beta_W + \beta_H \cos\theta)^2}{\cos\theta + x_W}}  
\right] \right\}
\eeq
with $\beta_{W,H}$ denoting the $W$, $H$ velocities in the c.m.\ frame,
and $x_W = (1- 2 M_H^2/\hat{s})/(\beta_W \beta_H)$. $\hat{s}^{1/2}$ is
the invariant energy of the $WW$ pair; $\theta$ is the Higgs 
production angle in the c.m.\ frame of $WW$. Integrating out the angular
dependence, the corresponding total cross section can be derived
\cite{djouadi} as
\beq
\hat{\sigma}_{LL} &=& \frac{G_F^2 M_W^4}{4\pi \hat{s}} \frac{\beta_H}
{\beta_W (1-\beta_W^2)^2} \Bigg\{ (1+\beta_W^2)^2 \left[1 + 
\frac{\lambda_{HHH}}
{(\hat{s} -M_H^2)/M_Z^2} \right]^2 \non\\
& + &\frac{16}{(1+\beta_H^2)^2-4\beta_H^2 \beta_W^2} 
\left[ \beta_H^2(
-\beta_H^2 x_W^2+4\beta_W \beta_H x_W -4\beta_W^2)
 + (1+\beta_W^2 -\beta_W^4)^2 \right] \non \\
& + &\frac{1}{\beta_W^2 \beta_H^2} \left( l_W +
 \frac{2x_W} {x_W^2-1} \right) 
 \left[ \beta_H ( \beta_H x_W-4\beta_W) (1+ \beta_W^2-
\beta_W^4 +3 x_W^2 \beta_H^2) \right. \non\\ 
&& \left. \hspace{4.2cm} + \beta_H^2 x_W 
(1-\beta_W^4 +13 \beta_W^2) 
- \frac{1}{x_W}(1+\beta_W^2-\beta_W^4)^2 \right] \non\\
& + & \frac{2(1+\beta_W^2)}{ \beta_W \beta_H} 
\left[1 + \frac{\lambda_{HHH}}{(\hat{s} -M_H^2)/M_Z^2} \right] 
\left[ l_W ( 1+\beta_W^2-\beta_W^4 -2\beta_W \beta_H x_W +
\beta_H^2 x_W^2) \right.
\non\\
& & \left. \hspace{5.75cm} +2\beta_H (x_W \beta_H  -2\beta_W ) \right] 
\Bigg\} 
\eeq
with $ l_W = \log [(x_W-1)/(x_W+1)]$. After folding the 
cross section of the subprocess with the longitudinal $W_L$ spectra 
\cite{repko},
\beq
f_L(z) = \frac{G_F M_W^2}{2\sqrt{2} \pi^2} \frac{1-z}{z}
\qquad [z = E_W/E_e]
\eeq
a rough estimate of the total $e^+ e^-$ cross section can be obtained;
it exceeds the exact value by about a factor 2 to 5 depending on the
collider energy and the Higgs mass. The estimate is useful
nevertheless for a transparent interpretation of the exact results.\s
\begin{table}
\begin{center}$
\begin{array}{|rl|rl||c|c|}\hline
\multicolumn{4}{|c||}{\sigma\;[\mathrm{fb}]} & WW & ZZ \str \\
\hline \hline
\sqrt{s}\!=\!\!\!\!&1\;\mathrm{TeV}& 
M_H\!=\!\!\!\!&110\;\mathrm{GeV} &
0.104 & 0.013 \str \\ 
\phantom{val} & \phantom{val} & \phantom{val} & 150\;\mathrm{GeV} &
0.042 & 0.006 \str \\
\phantom{val} & \phantom{val} & \phantom{val} & 190\;\mathrm{GeV} &
0.017 & 0.002 \str \\ \hline
\sqrt{s}\!=\!\!\!\!&1.6\;\mathrm{TeV}& 
M_H\!=\!\!\!\!&110\;\mathrm{GeV} &
0.334 &  0.043\str \\ 
\phantom{val} & \phantom{val} & \phantom{val} & 150\;\mathrm{GeV} &
0.183 & 0.024 \str \\
\phantom{val} & \phantom{val} & \phantom{val} & 190\;\mathrm{GeV} &
0.103 & 0.013 \str \\ \hline
\end{array}$
\end{center}
\caption{Total cross sections for SM pair production in $WW$ and $ZZ$ fusion at \ee colliders for two characteristic energies and masses in the intermediate range (unpolarized beams).}
\label{tab:SM}
\end{table}

For large $WW$ energies the process $WW \lra HH$ is dominated by
t-channel $W$ exchange which is independent of the trilinear Higgs
coupling. However, even at high c.m. energies the convoluted process
\ee $\lra \bar{\nu}_e \nu_e HH$ receives most of the contributions
from the lower end of the $WW$ energy spectrum so that the sensitivity
on $\lambda_{HHH}$ is preserved also in this domain. \s

The exact cross sections for off-shell $W$ bosons, transverse degrees
of freedom included, have been calculated numerically, based on the
semi-analytical CompHEP program~\cite{boos}. Electron and positron beams
are assumed to be polarized, giving rise to a cross section four times
larger than for unpolarized beams. The results are shown in Fig.~7a
for the three energies discussed before: $\sqrt{s}=500$~GeV, 1~TeV and
1.6~TeV. As expected, the fusion cross sections increase with rising
energy.  Again, the variation of the cross section with
$\kappa\lambda_{HHH}$, $\kappa = -1$ to $+2$, is demonstrated in
Fig.~7b for $\sqrt{s}=1$~TeV and $M_H=110$~GeV, and by the vertical
arrows for $\kappa=1/2$ to 3/2 in Fig.~7a. Due to the destructive
interference with the gauge part of the amplitude, the cross sections
drop with rising $\lambda_{HHH}$. The $ZZ$ fusion cross section is an
order of magnitude smaller than the $WW$ fusion cross section since
the $Z$ couplings of the electron/positron are small,
cf.~Table~\ref{tab:SM}.\s
\begin{figure}
\begin{center}
\includegraphics{sm-WW-HH-1.eps} \\[1cm]
\end{center}
Figure 7a: {\it The total cross section for WW double-Higgs fusion in 
the SM at three collider energies: $500$~GeV, $1$~TeV and $1.6$~TeV. 
The vertical arrows correspond to a variation of the trilinear Higgs 
coupling from $1/2$ to $3/2$ of the SM value.}
\label{fig:SM2}
\end{figure}
\begin{figure}
\begin{center}
\includegraphics{sm-WW-HH-2.eps} \\[1cm]
\end{center}
Figure 7b: {\it Variation of the cross section 
$\sigma(e^+ e^- \lra \bar{\nu}_e \nu_e HH)$ with the modified 
trilinear coupling $\kappa \lambda_{HHH}$ at a collider energy of 
$\sqrt{s}=1$~TeV and $M_H=110$~GeV.}
\label{fig:wwvar1}
\end{figure}
\setcounter{figure}{7}

It is apparent from the preceding discussion that double
Higgs-strahlung $e^+ e^- \to ZHH$ at moderate energies and $WW$ fusion
at TeV energies are the preferred channels for measurements of the
trilinear self-coupling $\lambda_{HHH}$ of the SM Higgs boson.
Electron and positron beam polarization enhance the cross sections by
factors 2 and 4 for Higgs-strahlung and $WW$ fusion, respectively.
Since the cross sections are small, high luminosity of the \ee ~linear
collider is essential for performing these fundamental experiments.
Even though the rates of order $10^3$ to $3 \cdot 10^3$ events for an
integrated luminosity of 2~ab$^{-1}$ as foreseen for TESLA, are
moderate, clear multi-$b$ signatures like $e^+ e^- \to
Z(b\bar{b})(b\bar{b})$ and $e^+ e^- \to(b\bar{b})(b\bar{b}) +
\fmslash{E}$ will help to isolate the signal from the background.\s

The complete reconstruction of the Higgs potential in the Standard
Model requires the measurement of the quadrilinear coupling
$\lambda_{HHHH}$, too. This coupling is sup\-pres\-sed relative to the
trilinear coupling effectively by a factor of the order of the weak
gauge coupling for masses in the lower part of the intermediate Higgs
mass range. The quadrilinear coupling can be accessed directly only
through the production of three Higgs bosons: \ee $\lra ZHHH$ and \ee
$\lra \bar{\nu}_e \nu_e HHH$. However, these cross sections are
reduced by three orders of magnitude compared to the corresponding
double-Higgs channels. As argued before, the signal amplitude
involving the four-Higgs coupling [as well as the irreducible
Higgs-strahlung amplitudes] is suppressed, leading to a reduction of
the signal cross section by a factor
$[\lambda_{HHHH}^2\lambda_0^4/16\pi^2]/[\lambda_{HHH}^2\lambda_0^2/M_Z^2]
\sim 10^{-3}$. Irreducible background diagrams are similarly
suppressed. Moreover, the phase space is reduced by the additional
heavy particle in the final state. A few illustrative examples for
triple Higgs-strahlung are listed in Table~\ref{tab:quadri}.

\begin{table}
\begin{center}$
\begin{array}{|rlrl|lrl|}\hline
\multicolumn{4}{|c|}{\phantom{\sigma(HH)\;[\mathrm{ab}]}} &
\multicolumn{3}{|c|}{\sigma($\ee$\to ZHHH)[\mathrm{ab}]} 
\str \\
\hline 
\sqrt{s}\!=\!\!\!\!&1\;\mathrm{TeV}& 
M_H\!=\!\!\!\!&110\;\mathrm{GeV} &
\hspace{0.5cm} 0.44 & [0.41/\!\!\!\!& 0.46] \str \\ 
\phantom{val} & \phantom{val} & \phantom{val} & 150\;\mathrm{GeV} &
\hspace{0.5cm}0.34 & [0.32/\!\!\!\!& 0.36] \str \\ 
\phantom{val} & \phantom{val} & \phantom{val} & 190\;\mathrm{GeV} &
\hspace{0.5cm}0.19 & [0.18/\!\!\!\!& 0.20] \str \\ \hline
\sqrt{s}\!=\!\!\!\!&1.6\;\mathrm{TeV}& 
M_H\!=\!\!\!\!&110\;\mathrm{GeV} &
\hspace{0.5cm}0.30 & [0.29/\!\!\!\!& 0.32] \str \\ 
\phantom{val} & \phantom{val} & \phantom{val} & 150\;\mathrm{GeV} &
\hspace{0.5cm}0.36 & [0.34/\!\!\!\!& 0.39] \str \\ 
\phantom{val} & \phantom{val} & \phantom{val} & 190\;\mathrm{GeV} &
\hspace{0.5cm}0.39 & [0.36/\!\!\!\!& 0.43] \str \\ \hline
\end{array}$
\end{center}
\caption{Representative values for triple SM Higgs-strahlung 
(unpolarized beams). The sensitivity to the quadrilinear coupling is 
illustrated by the variation of the cross sections when 
$\lambda_{HHHH}$ is altered by factors $1/2$ and $3/2$, as indicated 
in the square brackets.}
\label{tab:quadri}
\end{table}

\subsection*{3. The Supersymmetric Higgs Sector}

A large ensemble of Higgs couplings are present in supersymmetric
theories. Even in the minimal realization MSSM, six different
trilinear couplings $hhh$, $Hhh$, $HHh$, $HHH$, $hAA$, $HAA$ are
generated among the neutral particles, and many more quadrilinear
couplings \cite{dubinin}. Since in major parts of the MSSM parameter
space the Higgs bosons $H$, $A$, $H^\pm$ are quite heavy, we will
focus primarily on the production of light neutral pairs $hh$, yet the
production of heavy Higgs bosons will also be discussed where
appropriate.  The channels in which trilinear Higgs couplings can be
probed in \ee ~collisions, have been cataloged in
Table~\ref{tab:coup}.\s

Barring the exceptional case of very light pseudoscalar $A$ states,
$\lambda_{Hhh}$ is the only trilinear coupling that may be measured in
resonance decays, $H\to hh$, while all the other couplings
must be accessed in continuum pair production. The relevant mechanisms
have been ca\-te\-go\-ri\-zed in Fig.~\ref{fig:graphs} for double
Higgs-strahlung, associated triple Higgs production and $WW$
double-Higgs fusion.\s

\subsubsection*{3.1 Double Higgs-strahlung}

The (unpolarized) cross section for double Higgs-strahlung, $e^+ e^-
\to Zhh$, is modified \cite{djouadi} (see also \cite{osland}) with
regard to the Standard Model by $H$,$A$ exchange diagrams,
cf.~Fig.~\ref{fig:graphs}:
\beq
\frac{d \sigma (e^+ e^- \to Zhh)}{d x_1 d x_2} &=& 
\frac{\sqrt{2} G_F^3 M_Z^6}{384 \pi^3 s} 
\frac{v_e^2 + a_e^2}{(1- \mu_Z)^2}\, {\cal Z}_{11}
\label{zhh1}
\eeq
with
\beq
{\cal Z}_{11} &=&  {\mathfrak a}^2 f_0 + 
\frac{{\mathfrak a}}{2} \left[ 
\frac{\sin^2 (\beta-\alpha) f_3}{y_1 + \mu_{1Z}} + 
\frac{\cos^2 (\beta-\alpha) f_3}{y_1 + \mu_{1A}} \right] \non \\ 
&& {} + \frac{\sin^4 (\beta-\alpha)}{4\mu_Z (y_1+\mu_{1Z})} \left[ 
\frac{f_1}{y_1+\mu_{1Z}} + \frac{f_2}{y_2+\mu_{1Z}} \right] \non 
\\
&& {} + \frac{\cos^4 (\beta-\alpha)}{4\mu_Z (y_1+\mu_{1A})} \left[ 
\frac{f_1}{y_1+\mu_{1A}} + \frac{f_2}{y_2+\mu_{1A}} \right] \non\\
&& {} + \frac{\sin^2 2(\beta-\alpha)}{8\mu_Z (y_1+\mu_{1A})} \left[ 
\frac{f_1}{y_1+\mu_{1Z}} + \frac{f_2}{y_2+\mu_{1Z}} \right]
+ \Bigg\{ y_1 \leftrightarrow y_2 \Bigg\}
\label{zhh2}
\eeq
and
\beq
{\mathfrak a} = \left[ 
\frac{\lambda_{hhh}\sin(\beta-\alpha)}{y_3-\mu_{1Z}}
+ \frac{\lambda_{Hhh}\cos(\beta-\alpha)}{y_3 - \mu_{2Z}} \right] 
+ \frac{2 \sin^2(\beta-\alpha)}{y_1+\mu_{1Z}} 
+ \frac{2 \sin^2(\beta-\alpha)}{y_2+\mu_{1Z}} 
+ \frac{1}{\mu_Z}
\label{zhh3}
\eeq
[The notation follows the Standard Model, with $\mu_1=M_h^2/s$ and
$\mu_2=M_H^2/s$.] In parameter ranges in which the heavy neutral Higgs
boson $H$ or the pseudoscalar Higgs boson $A$ becomes resonant, the
decay widths are implicitly included by shifting the masses to complex
values $M \to M - i\Gamma/2$, {\it i.e.} $\mu_i \to \mu_i - i
\gamma_i$ with the reduced width $\gamma_i = M_i\Gamma_i/s$, and
by changing products of propagators $\pi_1 \pi_2$ to Re$(\pi_1
\pi_2^*)$.\s
\begin{figure}
\begin{center}
\epsfig{figure=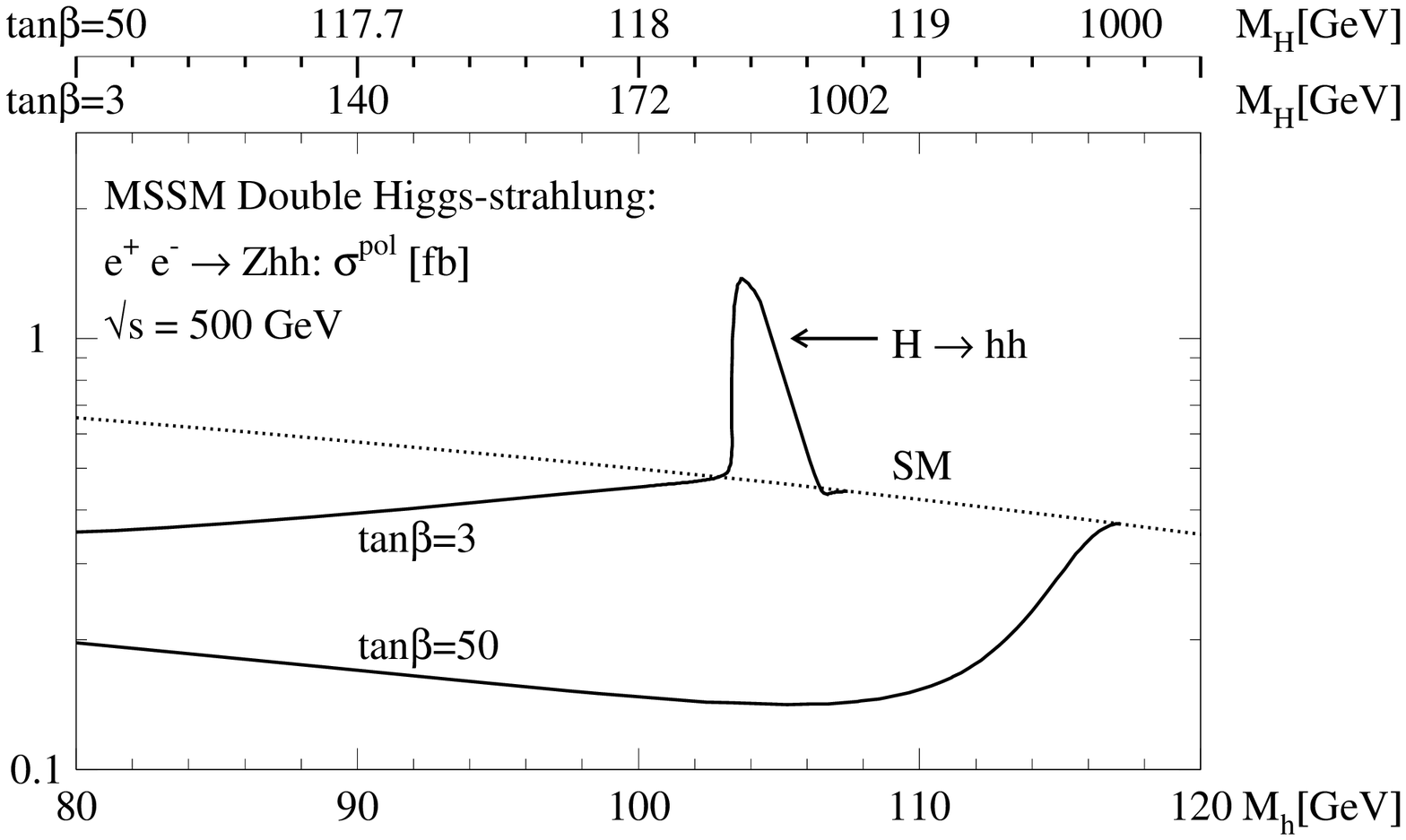,width=13cm}
\caption{
  Total cross sections for MSSM $hh$ production via double
  Higgs-strahlung at $e^+e^-$ linear colliders for $\tan\beta =3$, $50$ 
  and $\sqrt{s}=500\;\mathrm{GeV}$, including mixing effects 
  ($A = 1$~TeV, $\mu=-1/1$~TeV for $\tan\beta=3/50$). 
  The dotted line indicates the SM cross section.}
\label{fig:NLC/SUSY}
\end{center}
\end{figure}

The total cross sections are shown in Fig.~\ref{fig:NLC/SUSY} for the
$e^+ e^-$ collider energy $\sqrt{s}= 500$ GeV. The parameter $\tan
\beta$ is chosen to be 3 and 50 and the mixing parameters $A = 1$~TeV
and $\mu = -1$~TeV and $1$~TeV, respectively. If $\tan \beta$ and the
mass $M_h$ are fixed, the masses of the other heavy Higgs bosons are
predicted in the MSSM \cite{zerwas}. Since the vertices are suppressed
by $\sin/\cos$ functions of the mixing angles $\beta$ and $\alpha$,
the continuum $hh$ cross sections are suppressed compared to the
Standard Model. The size of the cross sections increases for moderate
$\tan \beta$ by nearly an order of magnitude if the $hh$ final state
can be generated in the chain $e^+ e^- \to ZH \to Zhh$ via resonant
$H$ Higgs-strahlung. If the light Higgs mass approaches the upper
limit for a given value of $\tan \beta$, the decoupling theorem drives
the cross section of the supersymmetric Higgs boson back to its
Standard Model value since the Higgs particles $A$, $H$, $H^\pm$
become asymptotically heavy in this limit. As a result of the
decoupling theorem, resonance production is not effective for large
tan$\beta$.  If the $H$ mass is large enough to allow decays to $hh$
pairs, the $ZZH$ coupling is already too small to generate a sizable
cross section.\s

\begin{figure}
\begin{center}
\epsfig{figure=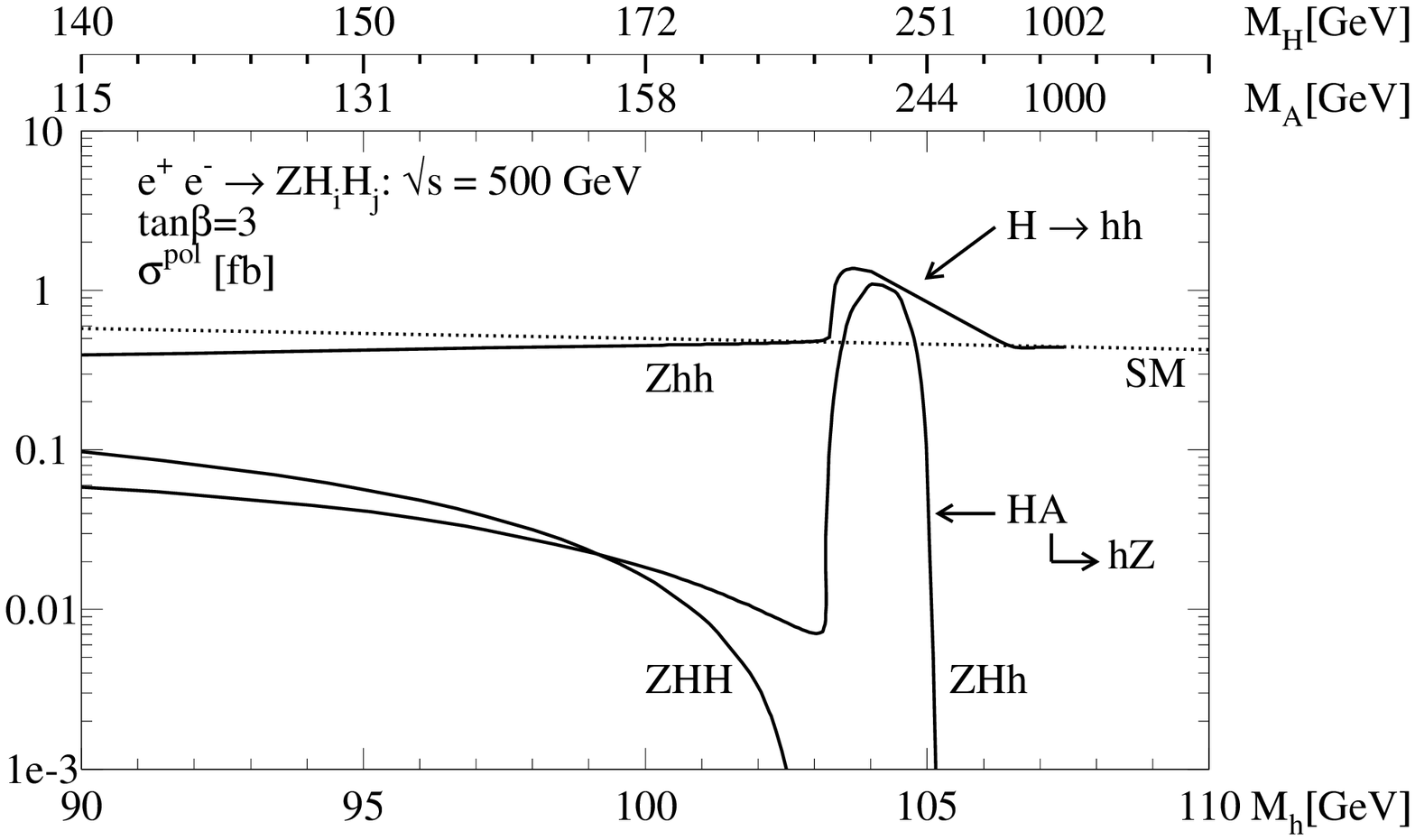,width=13cm}
\\
\end{center}
Figure 9a: {\it Cross sections for the processes $Zhh$, $ZHh$ and 
$ZHH$ for $\sqrt{s}=500$~GeV and tan$\beta=3$, including mixing 
effects ($A = 1$~TeV, $\mu=-1$~TeV).}
\end{figure}
The cross sections for other $ZH_i H_j$ [$H_{i,j}=h$, $H$] final
states are presented in the Appendix. While the basic structure
remains the same, the complexity increases due to unequal masses of
the final-state particles. The reduction of the $Zhh$ cross section is
partly compensated by the $ZHh$ and $ZHH$ cross sections so that their
sum adds up approximately to the SM value, as demonstrated in Fig.~9a
for tan $\beta=3$ at $\sqrt{s}=500$~GeV and $hh$, $Hh$ and $HH$ final
states. Evidently, if kinematically possible, the MSSM cross sections
add up to approximately the SM cross section.\pskip

\subsubsection*{3.2 Triple-Higgs Production}

The 2-particle processes \ee $\to ZH_i$ and \ee $\to AH_i$ are among
themselves and mutually complementary to each other in the MSSM
\cite{djoukal}, coming with the coefficients $\sin^2
(\beta-\alpha)/\cos^2(\beta-\alpha)$ and
$\cos^2(\beta-\alpha)/\sin^2(\beta-\alpha)$ for $H_i=h,$ $H$,
respectively. Since multi-Higgs final states are mediated by virtual
$h,$ $H$ bosons, the two types of self-complementarity and mutual
complementarity are also operative in double-Higgs production: \ee
$\to ZH_i H_j,$ $ZAA$ and $AH_i H_j,$ $AAA$. As the different
mechanisms are intertwined, the complementarity between these
3-particle final states is of more complex matrix form, as evident
from Fig.~\ref{fig:graphs}. \s

We will analyze in this section the processes involving only the light
neutral Higgs boson $h$, \ee $\to Ahh$, and three pseudoscalar Higgs
bosons $A$, \ee $\to AAA$. The more cumbersome expressions for heavy
neutral Higgs bosons $H$ are deferred to the Appendix. \s

In the first case one finds for the unpolarized cross section 
\beq
\frac{d\sigma [e^+ e^- \to Ahh]}{dx_1 dx_2} = 
\frac{G_F^3 M_Z^6}{768 \sqrt{2} \pi^3 s} 
\frac{v_e^2 + a_e^2}{(1-\mu_Z)^2} {\mathfrak A}_{11}
\eeq
where the function ${\mathfrak A}_{11}$ reads
\beq
{\mathfrak A}_{11} &=& \left[ 
\frac{c_1 \lambda_{hhh}} {y_3-\mu_{1A}}
+ \frac{c_2 \lambda_{H hh}} {y_3-\mu_{2A}} \right]^2 \frac{g_0}{2}
+ \frac{c_1^2 \lambda^2_{hAA}} {(y_1+\mu_{1A})^2} g_1
+ \frac{c_1^2 d_1^2 } {(y_1+\mu_{1Z})^2} g_2 
\non \\
&+& \left[ \frac{c_1 \lambda_{hhh} } {y_3-\mu_{1A}}
+ \frac{c_2 \lambda_{H hh}} {y_3-\mu_{2A}} \right]
\left[ \frac{c_1 \lambda_{hAA}} {y_1+\mu_{1A}} g_3
+ \frac{c_1 d_1 } {y_1+\mu_{1Z}} g_4 \right] \non \\
&+& 
\frac{c_1^2\lambda_{h AA}^2}{2(y_1+\mu_{1A})(y_2+\mu_{1A})}
g_5
+ \frac{c_1^2 d_1^2}{2(y_1+\mu_{1Z})(y_2+\mu_{1Z})}g_8 
\non \\
&+& \frac{c_1^2 d_1 \lambda_{h AA}}{(y_1+\mu_{1A})(y_1+\mu_{1Z})}g_6
+\frac{c_1^2 d_1 \lambda_{h AA}}{(y_1+\mu_{1A})(y_2+\mu_{1Z})}g_7
\non \\
&+& \Bigg\{ y_1 \leftrightarrow y_2 \Bigg\}
\eeq
with $\mu_{1,2} = M_{h,H}^2/s$ and the vertex coefficients
\beq
c_1/c_2 = \cos (\beta-\alpha)/-\sin(\beta-\alpha) \qquad \mathrm{and} 
\qquad d_1/d_2 = \sin (\beta - \alpha)/\cos(\beta-\alpha)
\eeq
The coefficients $g_k$ are given by
\beq
g_0 &=& \mu_Z[(y_1+y_2)^2-4\mu_A] \non \\
g_1 &=& \mu_Z[y_1^2-2y_1-4\mu_1+1] \non \\
g_2 &=& \mu_Z [y_1(y_1+2)+ 4y_2(y_2+y_1-1)+1- 4(\mu_1+2\mu_A)]
+ (\mu_1-\mu_A)^2 \non\\
&& [8+[(1-y_1)^2-4\mu_1]/\mu_Z] + (\mu_1-\mu_A) 
[4 y_2(1+y_1)+2(y_1^2-1)]\non\\
g_3 &=& 2\mu_Z(y_1^2- y_1+y_2+y_1 y_2-2\mu_A) \non \\
g_4 &=& 2\mu_Z(y_1^2+y_1+2y_2^2-y_2+3y_1 y_2 -6 
\mu_A) \non\\ 
&&+2(\mu_1-\mu_A)(y_1^2- y_1+ y_2  + y_1y_2 -2\mu_A) 
    \non \\
g_5 &=& 2\mu_Z(y_1+ y_2+y_1y_2+4\mu_1-2\mu_A-1) \non \\
g_6 &=& 2\mu_Z(y_1^2+ 2 y_1y_2+ 2y_2+4\mu_1-4\mu_A-1) \non \\
 && + 2(\mu_1-\mu_A)(y_1^2-2y_1-4\mu_1+1) \non \\
g_7 &=& 2[ \mu_Z(2y_1^2-3y_1+y_1y_2+y_2-4\mu_1-2\mu_A+1) 
\non \\
 && + (\mu_1-\mu_A)(y_1+y_1y_2+y_2 +4\mu_1-2\mu_A-1)] \non \\
g_8 &=& 2 \left\{ \mu_Z(y_1+y_2+2y_1^2+2y_2^2+5y_1 y_2 -1 + 4\mu_1 
 -10\mu_A)\right. \non \\
&& +4(\mu_1-\mu_A)(-2\mu_1-\mu_A-y_1-y_2+1) \non \\
&& + [2(\mu_1-\mu_A)((y_1+y_2+y_1y_2+y_1^2+y_2^2-1)\mu_Z
+2\mu_1^2+4\mu_A^2-\mu_1+\mu_A)
\non\\
&& + \left.
6\mu_A(\mu_A^2-\mu_1^2) + (\mu_1-\mu_A)^2 (1+y_1)(1+y_2)]/\mu_Z \right\}
\eeq
The notation of the kinematics is the same as for Higgs-strahlung. \s

Since only a few diagrams contribute to triple $A$ production,
cf.~Fig.~\ref{fig:graphs}, the expression for this cross section is
exceptionally simple:
\beq
\frac{d\sigma [e^+ e^- \to AAA]}{dx_1 dx_2} = 
\frac{G_F^3 M_Z^6}{768 \sqrt{2} \pi^3 s} 
\frac{v_e^2 + a_e^2}{(1-\mu_Z)^2} {\mathfrak A}_{33}
\eeq
where
\beq
{\mathfrak A}_{33} = D_3^2 g_0+ D_1^2 g_1 +D_2^2 g_1'
- D_3 D_1 g_3- D_3 D_2 g_3' + D_1 D_2 g_5
\eeq
and
\beq
D_k= \frac{\lambda_{hAA} c_1} {y_k-\mu_{1A}}
   + \frac{\lambda_{HAA} c_2} {y_k-\mu_{2A}} 
\eeq
The scaled mass parameter $\mu_1$ must be replaced by $\mu_A$ in the 
coefficients $g_i$ and $g_i'$ defined earlier.\s

\begin{figure}
\begin{center}
\epsfig{figure=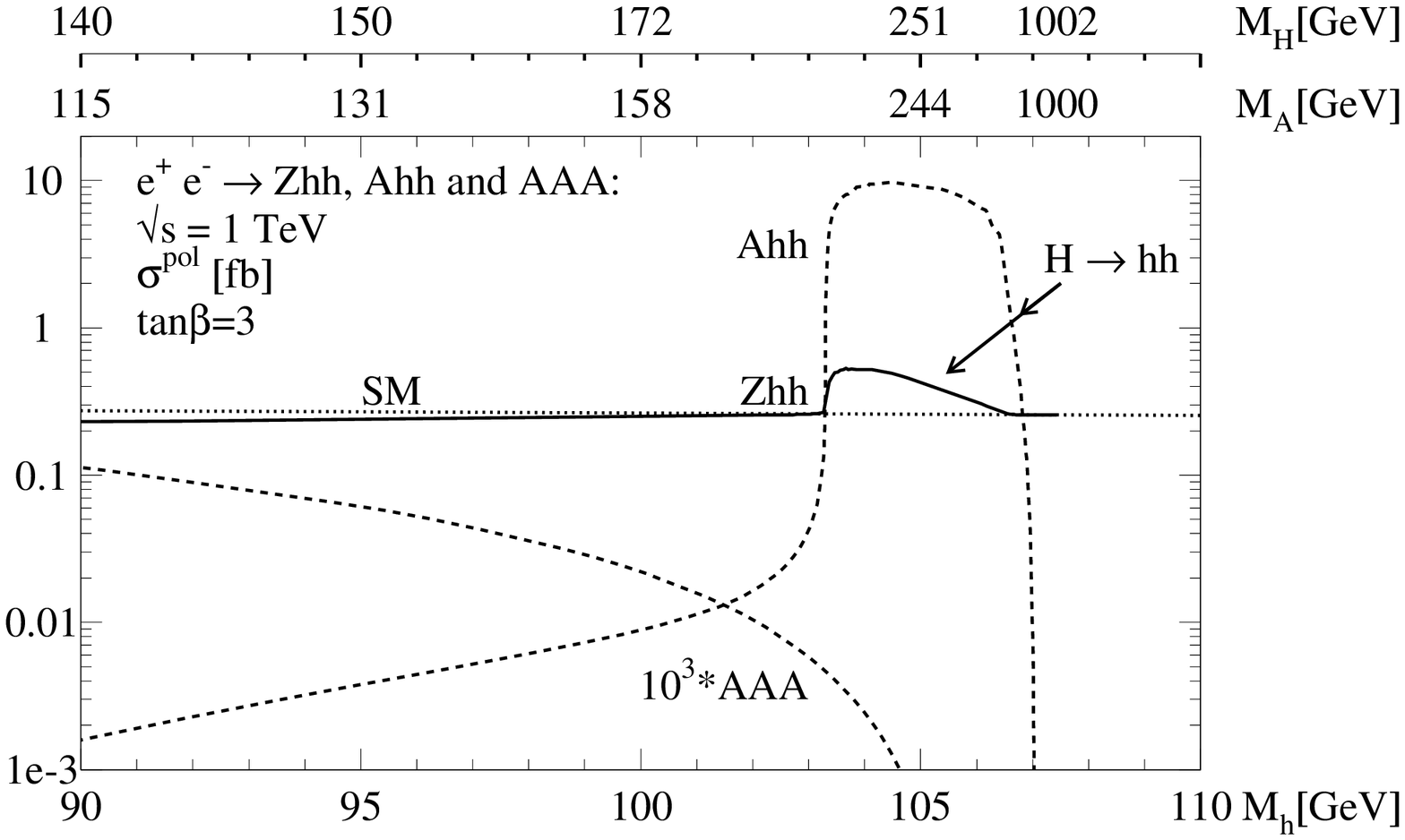,width=13cm}\\[0.9cm]
\end{center}
Figure 9b: {\it Cross sections of the processes Zhh, Ahh and AAA for 
$\tan\beta = 3$ and $\sqrt{s}=1$~TeV, including mixing effects 
($A=1$~TeV, $\mu=-1$~TeV.)}
\label{fig:ahh}
\end{figure}
\setcounter{figure}{9}
The size of the total cross section $\sigma(e^+ e^-\to Ahh)$ and
$\sigma(e^+ e^-\to AAA)$ is compared with double Higgs-strahlung
$\sigma (e^+ e^-\to Zhh)$ in Fig.~9b for tan $\beta = 3$ at
$\sqrt{s}= 1$~TeV. Both these cross sections involving pseudoscalar
Higgs bosons are small in the continuum. The effective coupling in the
chain $Ah_{virt} \to Ahh$ is $\cos(\beta-\alpha) \lambda_{hhh}$ while
in the chain $AH_{virt} \to Ahh$ it is $\sin(\beta -\alpha)
\lambda_{Hhh}$; both products are small either in the first or second
coefficient. Only for resonance $H$ decays $AH \to Ahh$ the cross
section becomes very large. A similar picture evolves for the triple
$A$ final state. The chain $Ah_{virt} \to AAA$ is proportional to the
coefficient $\cos(\beta-\alpha)\lambda_{hAA}$ in which one of the
terms is always small. The chain $AH_{virt} \to AAA$, on the other
hand, is proportional to $\sin(\beta-\alpha) \lambda_{HAA}$; for this
coefficient the trilinear coupling $\lambda_{HAA}$ is only of order
1/2 so that, together with phase space suppression, the cross section
remains small in the entire parameter space. \pskip

\subsubsection*{3.3 $WW$ Double-Higgs Fusion}
 
The $WW$ fusion mechanism for the production of supersymmetric Higgs
pairs can be treated in the same way. The dominant longitudinal
amplitude for on-shell $W$ bosons involves $A$, $H$ and $H^\pm$ exchange
diagrams in addition to the SM-type contributions:
\beq
{\cal M}_{LL} &=& \frac{G_F \hat{s}}{\sqrt{2}} \left\{ (1+\beta_W^2) 
\left[ 1 + 
\frac{\lambda_{hhh}\sin(\beta-\alpha)}{(\hat{s}-M_h^2)/M_Z^2} + 
\frac{\lambda_{Hhh}\cos(\beta-\alpha)}{(\hat{s}-M_H^2)/M_Z^2} \right] 
\right. \non\\
&+& \frac{\sin^2(\beta-\alpha)}{\beta_W \beta_h} \left[ 
\frac{(1-\beta_W^4) + 
(\beta_W - \beta_h \cos\theta)^2}{\cos\theta-x_W} - 
\frac{(1-\beta_W^4) + 
(\beta_W + \beta_h \cos\theta)^2}{\cos\theta+x_W} \right]
\non \\
&+& \left. \frac{\cos^2(\beta-\alpha)}{\beta_W \beta_h} \left[ 
\frac{(\beta_W - \beta_h \cos\theta)^2}{\cos\theta-x_+} - 
\frac{(\beta_W + \beta_h \cos\theta)^2}{\cos\theta+x_+} \right] 
\right\}
\eeq
As before, $\hat{s}^{1/2}$ is the c.m.\ energy of the subprocess,
$\theta$ the scattering angle, $\beta_W$ and $\beta_h$ are the
velocities of the $W$ and $h$ bosons, and
\begin{equation}
x_W = \frac{1-2\mu_h}{\beta_W \beta_h} \qquad {\rm and} \qquad
x_+ = \frac{1-2\mu_h+2\mu_{H^\pm}-2\mu_W}{\beta_W\beta_h} 
\end{equation}
After integrating out the angular dependence, the total cross section
of the fusion subprocess is given by the expression
\beq 
\hat{\sigma}_{LL} &=&
\frac{G_F^2 M_W^4}{4\pi \hat{s}} \frac{\beta_h}{ \beta_W
  (1-\beta_W^2)^2} \left\{ (1+\beta_W^2)^2 \left[ \frac{
      \lambda_{hhh} d_1}{(\hat{s} -M_h^2)/M_Z^2} + \frac{
      \lambda_{Hhh} d_2}{(\hat{s} -M_H^2)/M_Z^2} + 1\right]^2 \right. 
    \non \\
&& {}+ \frac{2(1+\beta_W^2)} {\beta_W \beta_h } \left[
  \frac{\lambda_{hhh}d_1}{(\hat{s} -M_h^2)/M_Z^2} + \frac{
    \lambda_{Hhh}d_2}{(\hat{s} -M_H^2)/M_Z^2} + 1 \right] \left[
  d_1^2 a_1^W  + c_1^2  a_1^+   \right] \non \\
&& {}+ \left. \left( \frac{d_1^2}{\beta_W \beta_h } \right)^2 a_2^W +
  \left( \frac{c_1^2}{\beta_W \beta_h } \right)^2 a_2^+ + 4
  \left(\frac{c_1^2 d_1^2}{\beta_W^2 \beta_h^2 } \right)
  [a_3^W + a_3^+ ] \right\} 
\eeq 
with 
\beq 
a_1^W &=& [ (x_W \beta_h
-\beta_W)^2 + r_W ] l_W +2 \beta_h (x_W \beta_h -2\beta_W ) \non \\
a_2^W &=& \left[ \frac{1}{x_W} l_W + \frac{2} {x_W^2-1} \right] \bigg[
x_W^2 \beta_h^2 (3 \beta_h^2 x_W^2
+2 r_W +14 \beta_W^2) \non \\
&& {}-(\beta_W^2+ r_W)^2 -4 \beta_h \beta_W x_W
(3 \beta_h^2 x_W^2 +\beta_W^2+r_W ) \bigg] \non \\
&& {}- \frac{4}{x_W^2-1} \left[ \beta_h^2 (\beta_h^2 x_W^2
  +4 \beta^2_W - 4 \beta_h x_W \beta_W) - (\beta_W^2 +r_W)^2 \right] 
\non \\
a_3^W&=& \frac{1}{x_+^2 - x_W^2} \, l_W \bigg[ 2 \beta_W \beta_h x_W
[(\beta_W^2+x_W^2 \beta_h^2)(x_W+x_+)
+ x_W r_W +x_+ r_+]  \non \\
&& {}-x_+( r_+ + r_W + \beta_h^2 x_W^2)(\beta_W^2 + \beta_h^2 x_W^2)
-\beta_W^2( x_+ \beta_W^2 +4\beta_h^2 x_W^3+
x_+ x_W^2 \beta_h^2 ) \non \\
&& {}- x_+ r_W r_+ \bigg] + \beta_h^2 \left[ 
\beta_h^2 x_+ x_W -2 \beta_W
  \beta_h (x_W+x_+) + 4\beta_W^2 \right] \non \\
a_i ^+ &\equiv& a_i^W \ 
(x_W \leftrightarrow x_+ \ , \ r_W \leftrightarrow r_+ )
\eeq 
and $r_W = 1- \beta^4_W$, $r_+ = 0$. \s

The final cross sections have been calculated for off-shell $W$'s and
transverse polarizations included, i.e. without relying on the $LL$ and
the equivalent $W$-boson approxi\-ma\-tion. The \ee ~beams are assumed
to be polarized. For modest $\tan\beta$, the $hh$ continuum production
is slightly suppressed by the mixing coefficients with regard to the
Standard Model, Fig.~10a. The cross section is strongly
enhanced in the parameter range where the fusion subprocess is
resonant, $WW\to H \to hh$.  For large $\tan\beta$ the $WW$ fusion
cross section is strongly suppressed by one to two orders of magnitude
and resonance decay is not possible any more. This is a consequence of
the small gauge couplings in this parameter range which are
drastically reduced by the mixing coefficients.  Since the second
CP-even Higgs boson $H$ is fairly light for these parameters, the
small $hh$ continuum production is complemented by $Hh$ and $HH$
production channels, as evident from Fig.~10b.  The cross sections for
the production of $Hh$, $HH$ and $AA$ pairs are cataloged in the
Appendix. \pskip
\begin{figure}
\begin{center}
\includegraphics{susy-mix-WW-HH-1.eps}\\[1cm]
\end{center}
Figure 10a: {\it Total cross sections for MSSM $hh$ production via 
double WW double-Higgs fusion at $e^+e^-$ linear colliders for 
$\tan\beta = 3,$ $50$ and $\sqrt{s}=1.6$~TeV, including mixing effects 
($A = 1$~TeV, $\mu=-1/1$~TeV for $\tan\beta=3/50$).}
\label{fig:WW/SUSY}
\end{figure}
\begin{figure}
\begin{center}
\includegraphics{susy-mix-WW-HH-2.eps} \\[1cm]
\end{center}
Figure 10b: {\it Total cross sections for WW double-Higgs fusion with 
$hh$, $Hh$ and $HH$ final states for $\sqrt{s}=1.6$~TeV and 
tan$\beta = 50$, including mixing effects ($A = 1$~TeV, $\mu=1$~TeV).}
\end{figure}
\setcounter{figure}{10}

\subsubsection*{3.4 Sensitivity Areas}

The results obtained in the preceding sections can be summarized in
compact form by constructing sensitivity areas for the trilinear SUSY
Higgs couplings based on the cross sections for double Higgs-strahlung
and triple Higgs production. $WW$ double-Higgs fusion can provide
additional information on the Higgs self-couplings. \s

The sensitivity areas will be defined in the $[M_A,$ tan$\beta]$ plane
\cite{djouadi}. The criteria for accepting a point in the plane as
accessible for the measurement of a specific trilinear coupling
are set as follows:
\beq
\begin{array}{l l} 
(i) & \sigma [\lambda] > 0.01~{\rm fb} \\
(ii) & {\rm var}\{ \lambda \to (1\pm \frac{1}{2})\lambda \} > 
2~{\rm st.dev.}
\{ \lambda \} \quad {\rm for} \quad \int {\cal L} = 2~{\rm ab}^{-1}
\end{array} 
\eeq
The first criterion demands at least 20 events in a sample collected
for an integrated lu\-mi\-no\-si\-ty of 2~ab$^{-1}$, corresponding to
about the lifetime of a high-luminosity machine such as TESLA. The
second criterion demands a 50\% change of the signal parameter to
exceed a statistical fluctuation of 2 standard deviations. Even though
the two criteria may look quite loose, tightening $(i)$ and/or $(ii)$
does not have a large impact on the size of the sensitivity areas in
the $[M_A,$ tan$\beta]$ plane, see Ref.~\cite{osland}. For the sake of
simplicity, the \ee ~beams are assumed to be unpolarized and mixing
effects are neglected.\s

\begin{figure}
\begin{center}
\epsfig{figure=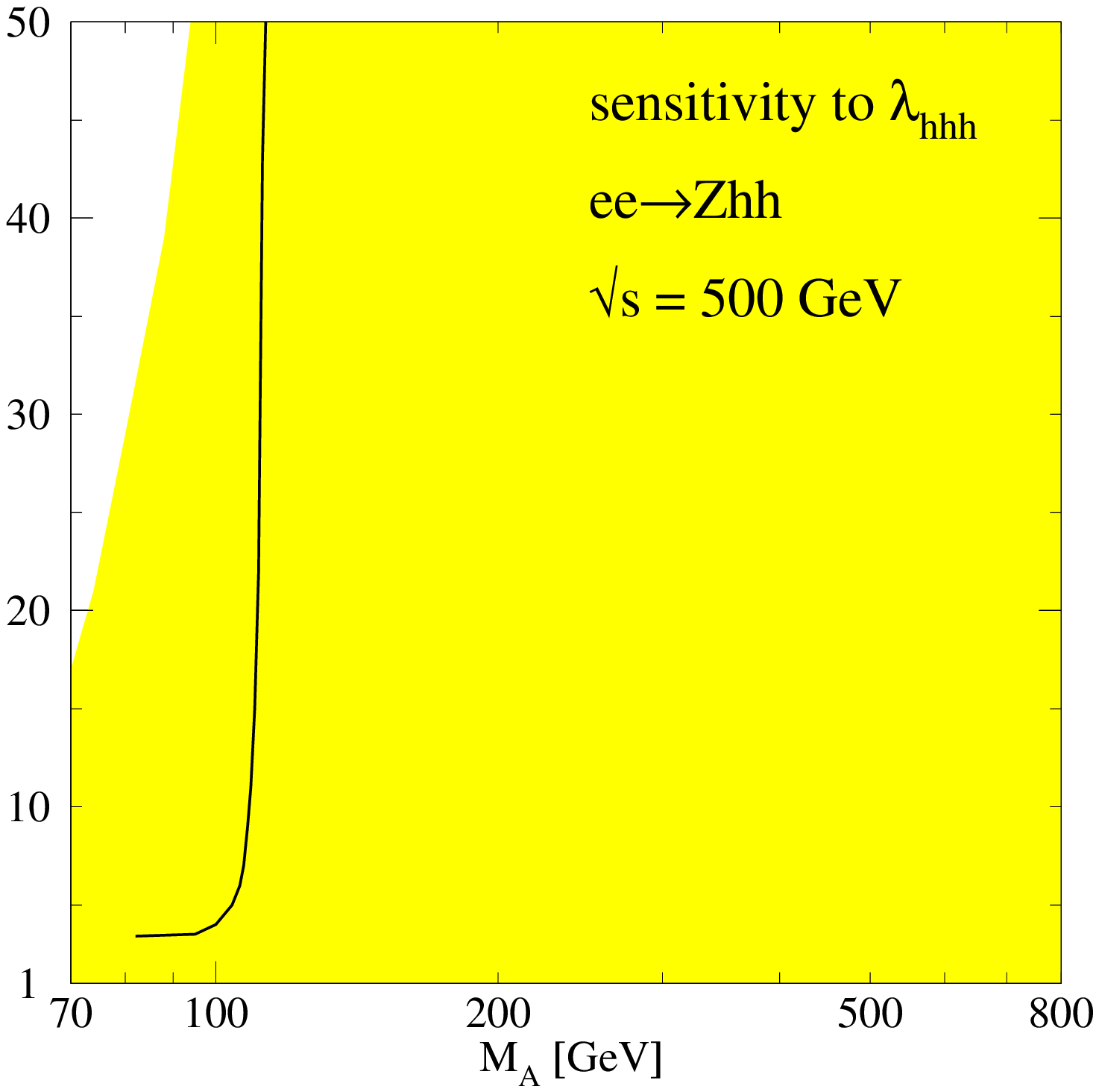,width=7cm}
\hspace{1cm}
\epsfig{figure=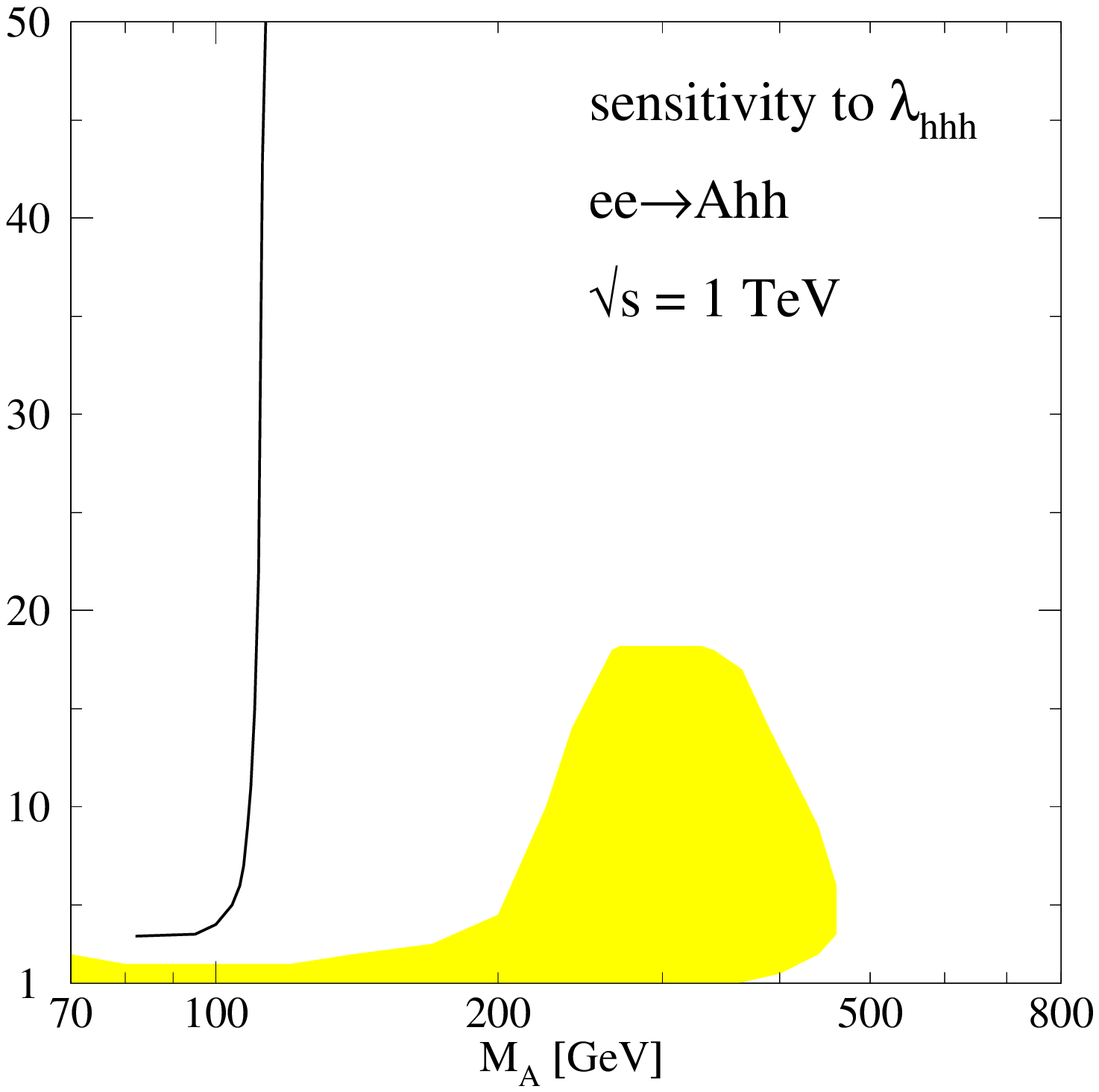,width=7cm}
\end{center}
\vspace{1.5cm}
\begin{center}
\epsfig{figure=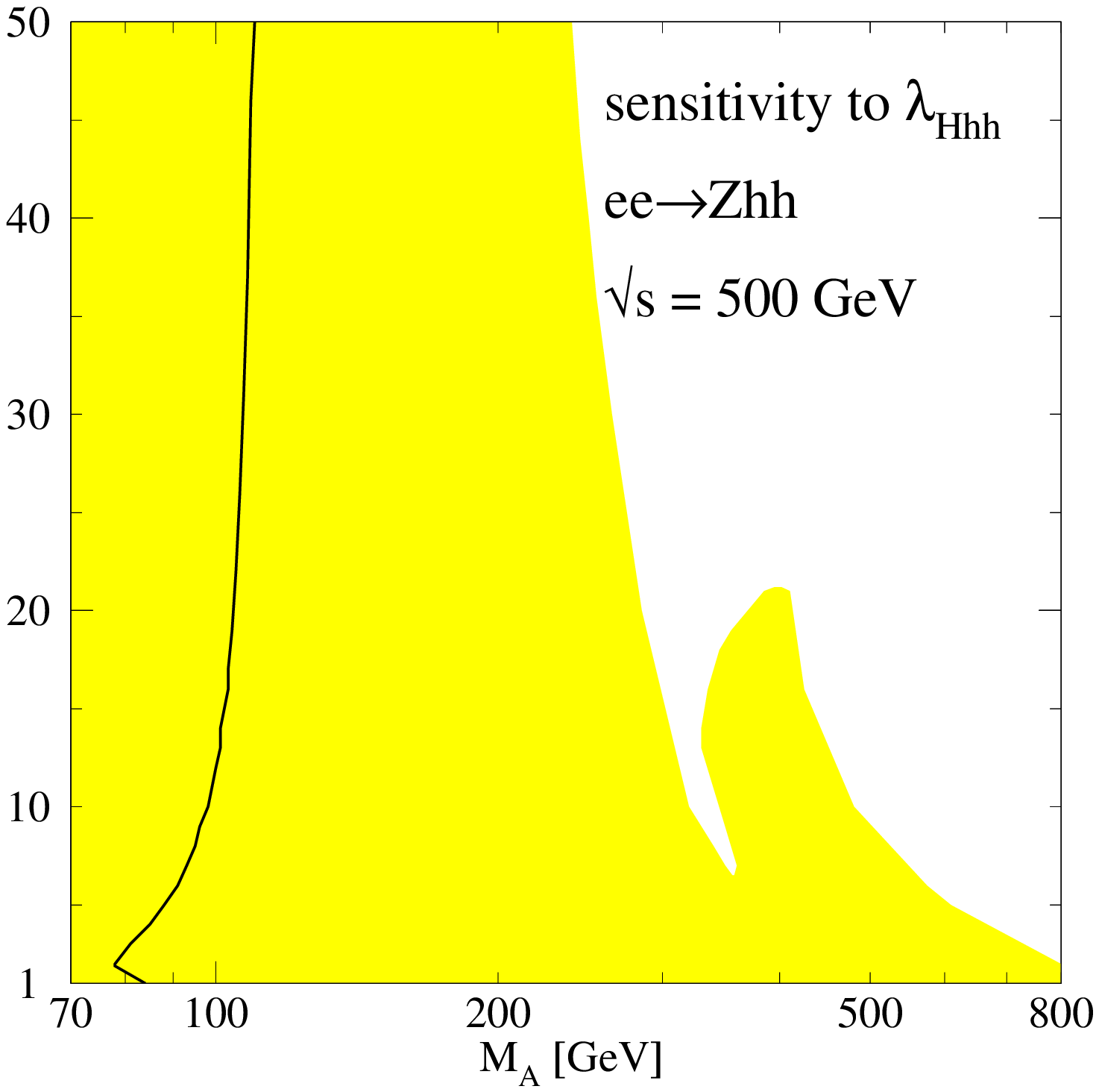,width=7cm}
\hspace{1cm}
\epsfig{figure=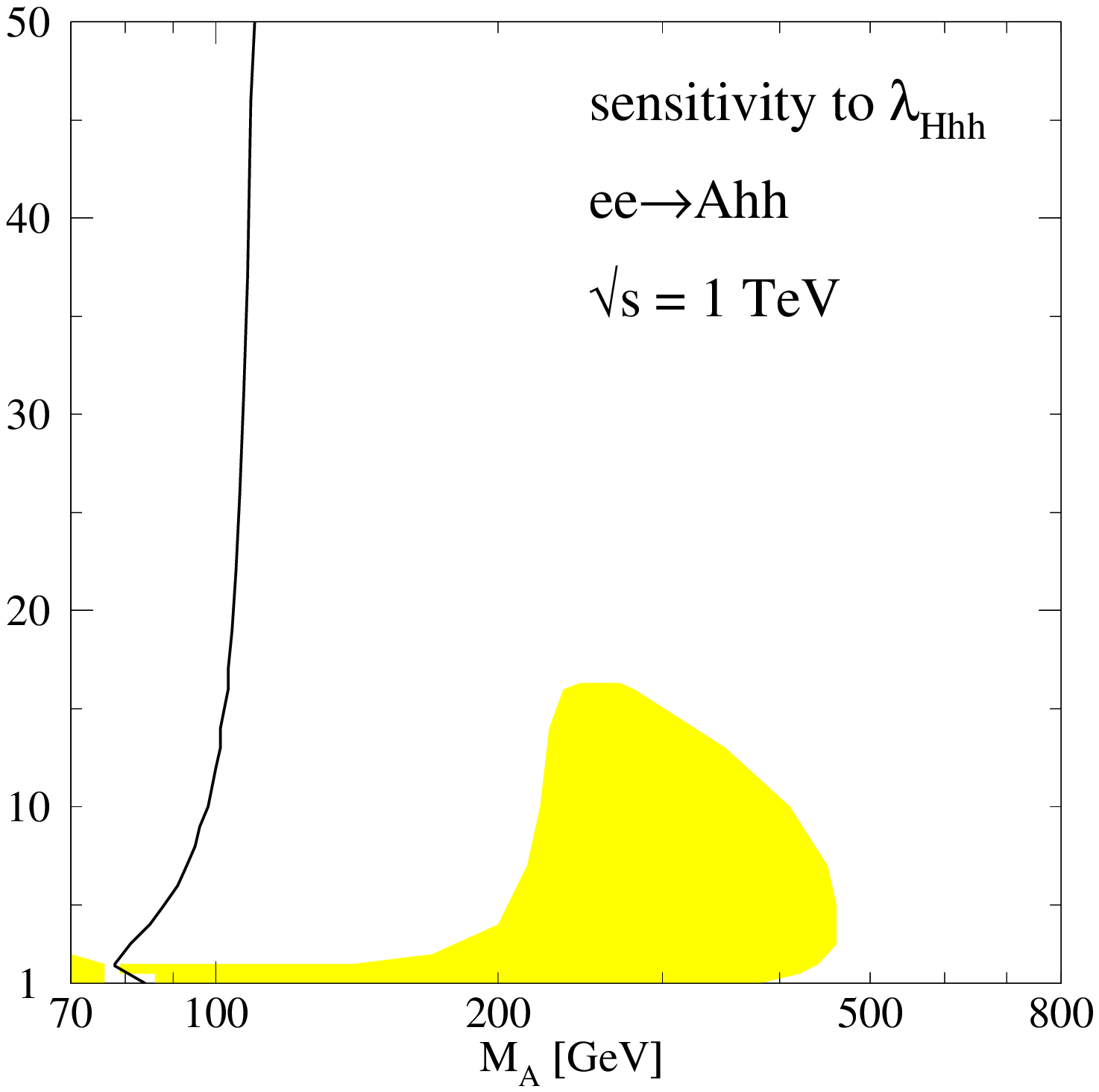,width=7cm}\\[0.5cm]
\caption{Sensitivity to $\lambda_{hhh}$ and $\lambda_{Hhh}$ in the 
processes \ee$\to Zhh$ and \ee$\to Ahh$ for collider energies 
$500$~GeV and $1$~TeV, respectively (no mixing). [Vanishing trilinear 
couplings are indicated by contour lines.]}
\label{fig:s1}
\end{center}
\end{figure}

\begin{figure}
\begin{center}
\epsfig{figure=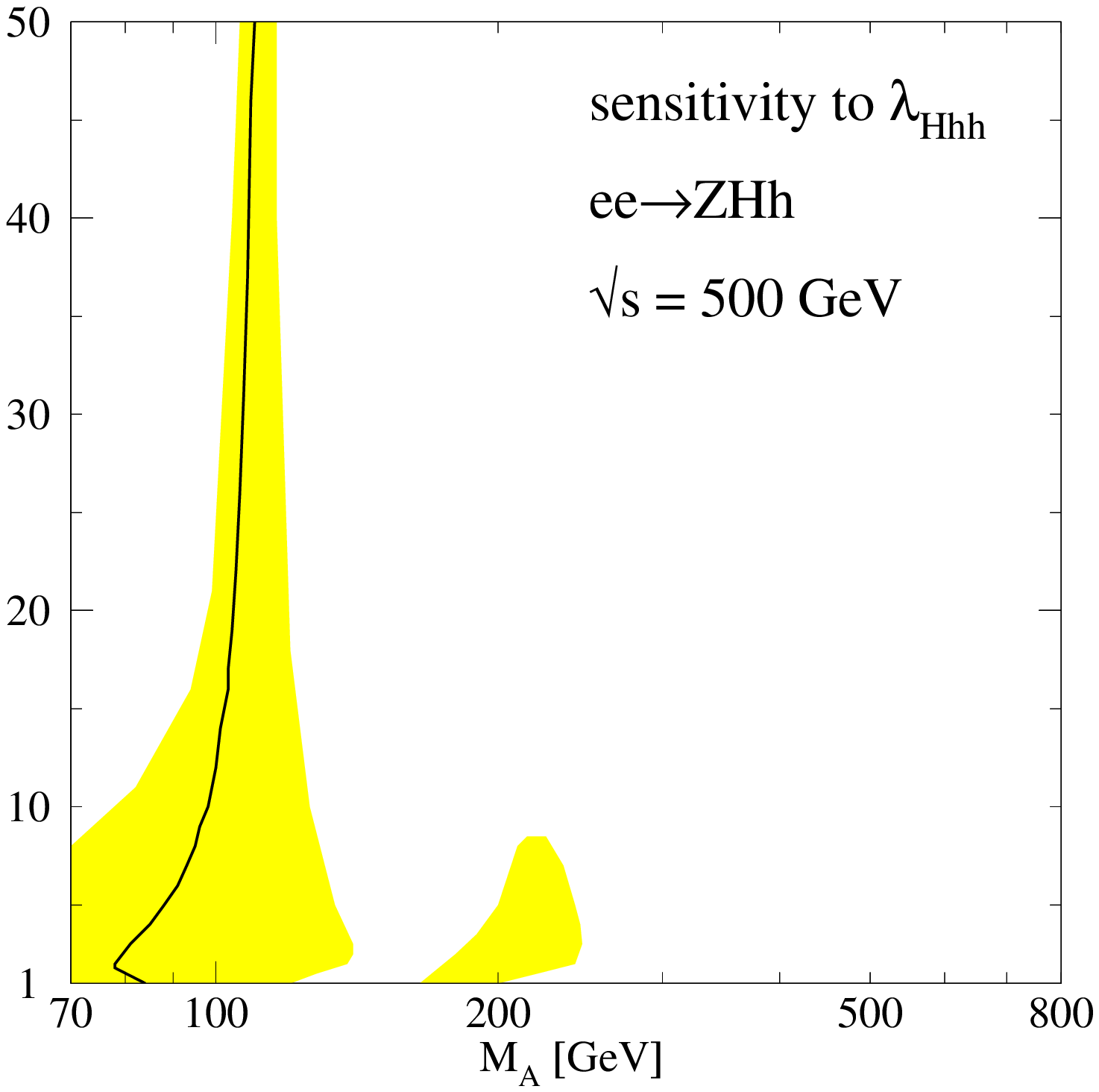,width=7cm}
\hspace{1cm}
\epsfig{figure=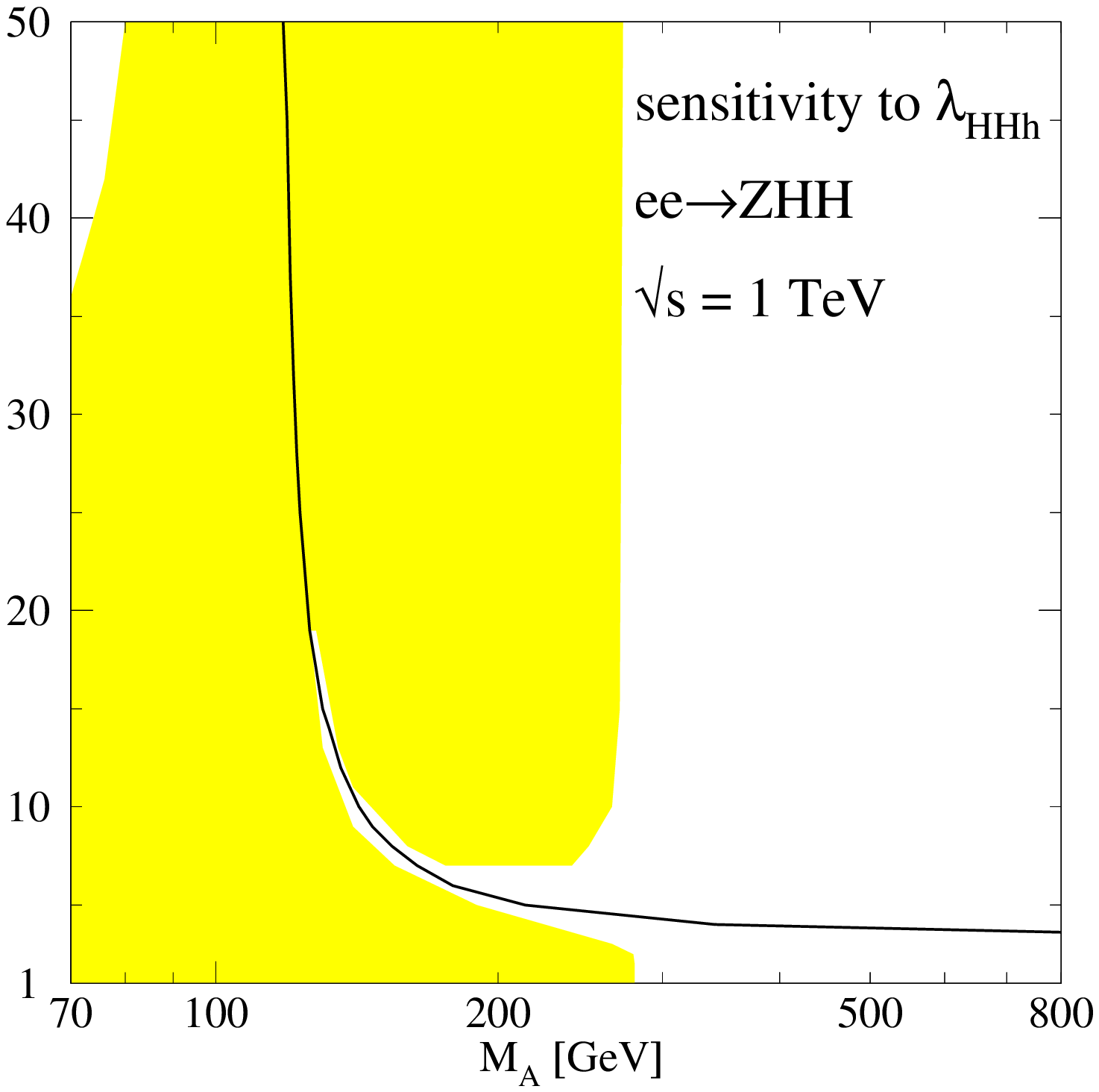,width=7cm}
\end{center}
\vspace{1.5cm}
\begin{center}
\epsfig{figure=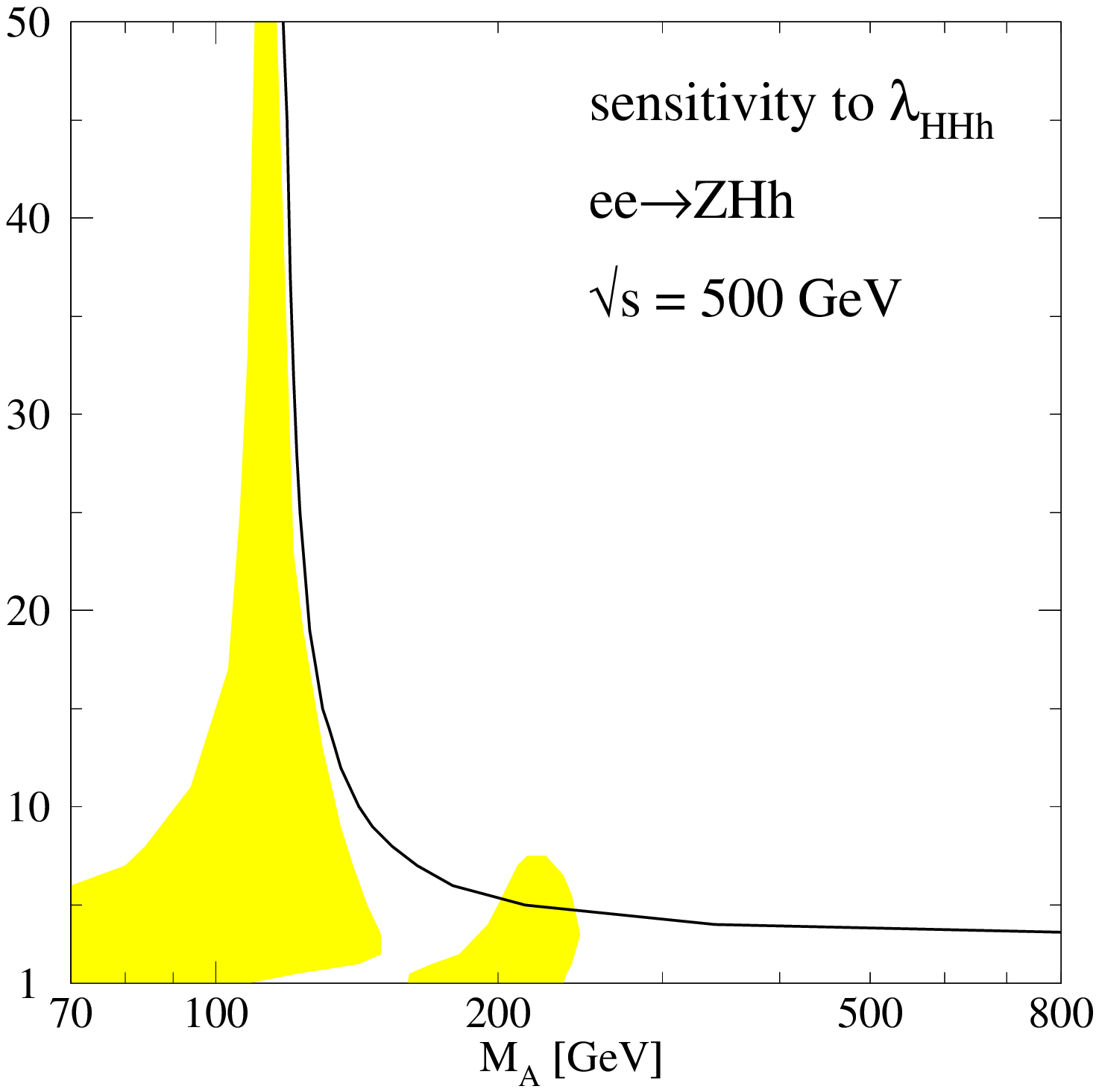,width=7cm}
\hspace{1cm}
\epsfig{figure=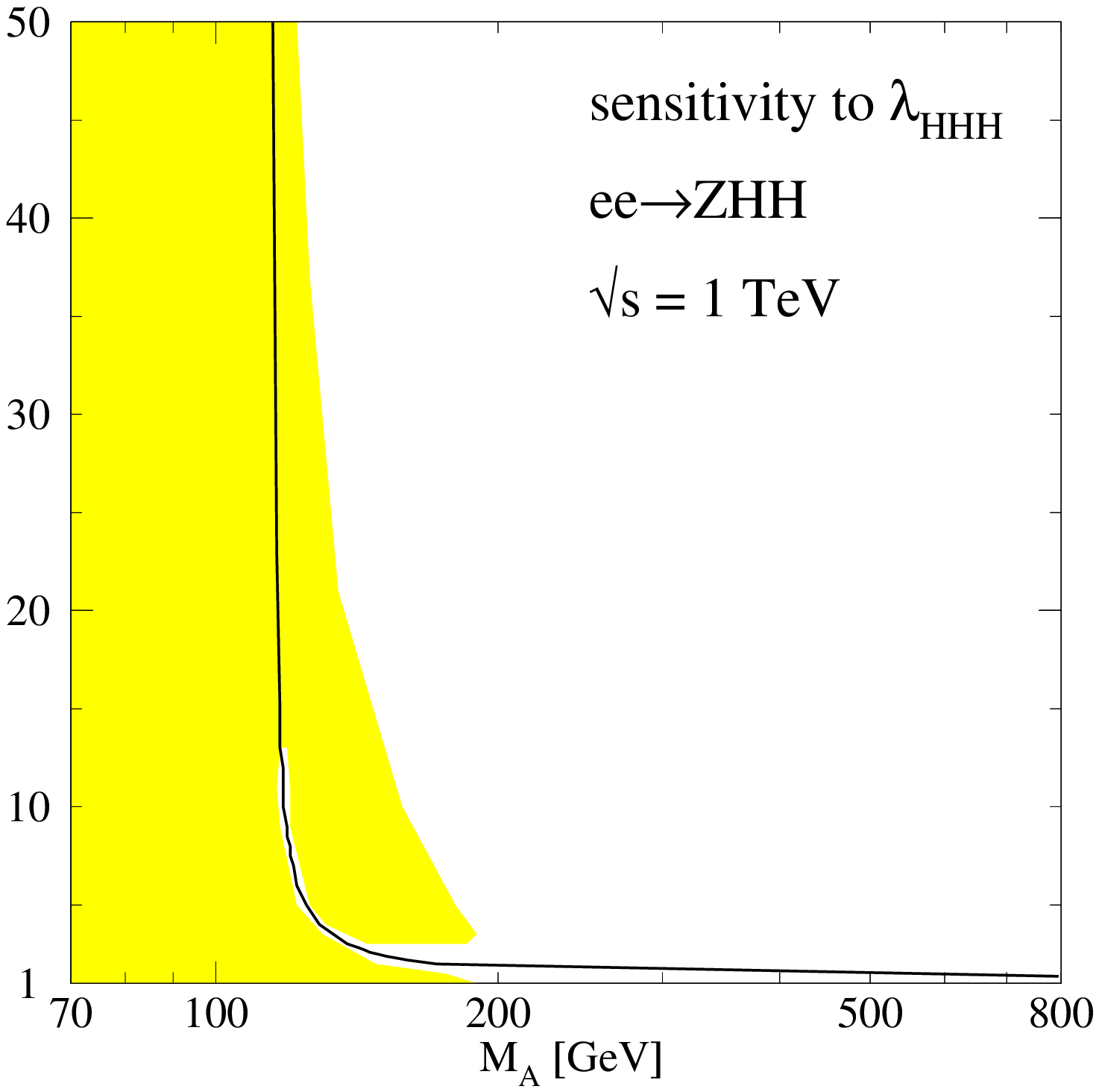,width=7cm}\\[0.5cm]
\caption{Sensitivity to $\lambda_{Hhh}$, $\lambda_{HHh}$ and 
$\lambda_{HHH}$ in the processes \ee$\to ZHh$ and \ee$\to ZHH$ 
for collider energies $500$~GeV and $1$~TeV, respectively (no mixing).}
\label{fig:s2}
\end{center}
\end{figure}

\begin{figure}
\begin{center}
\epsfig{figure=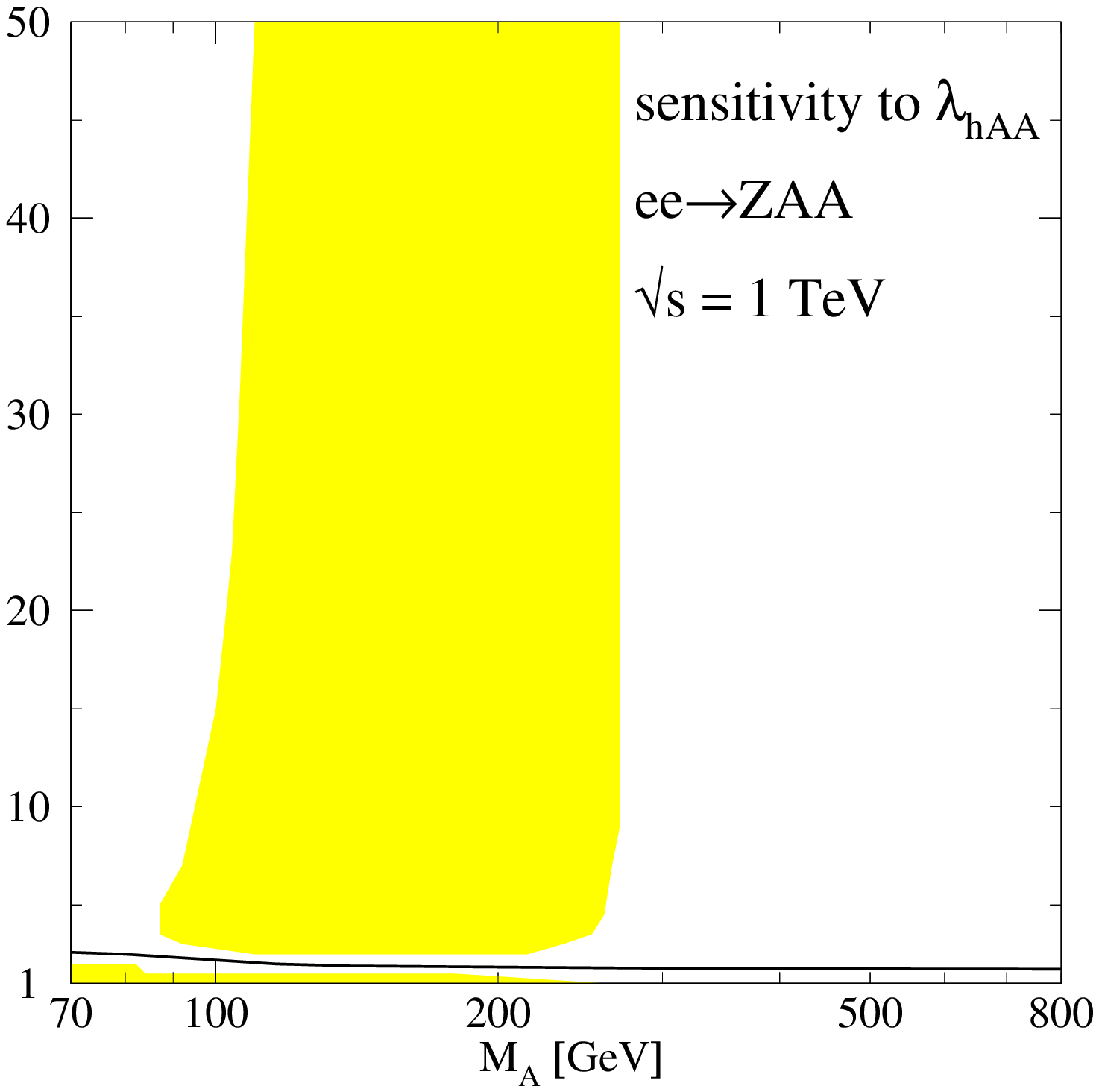,width=7cm}
\hspace{1cm}
\epsfig{figure=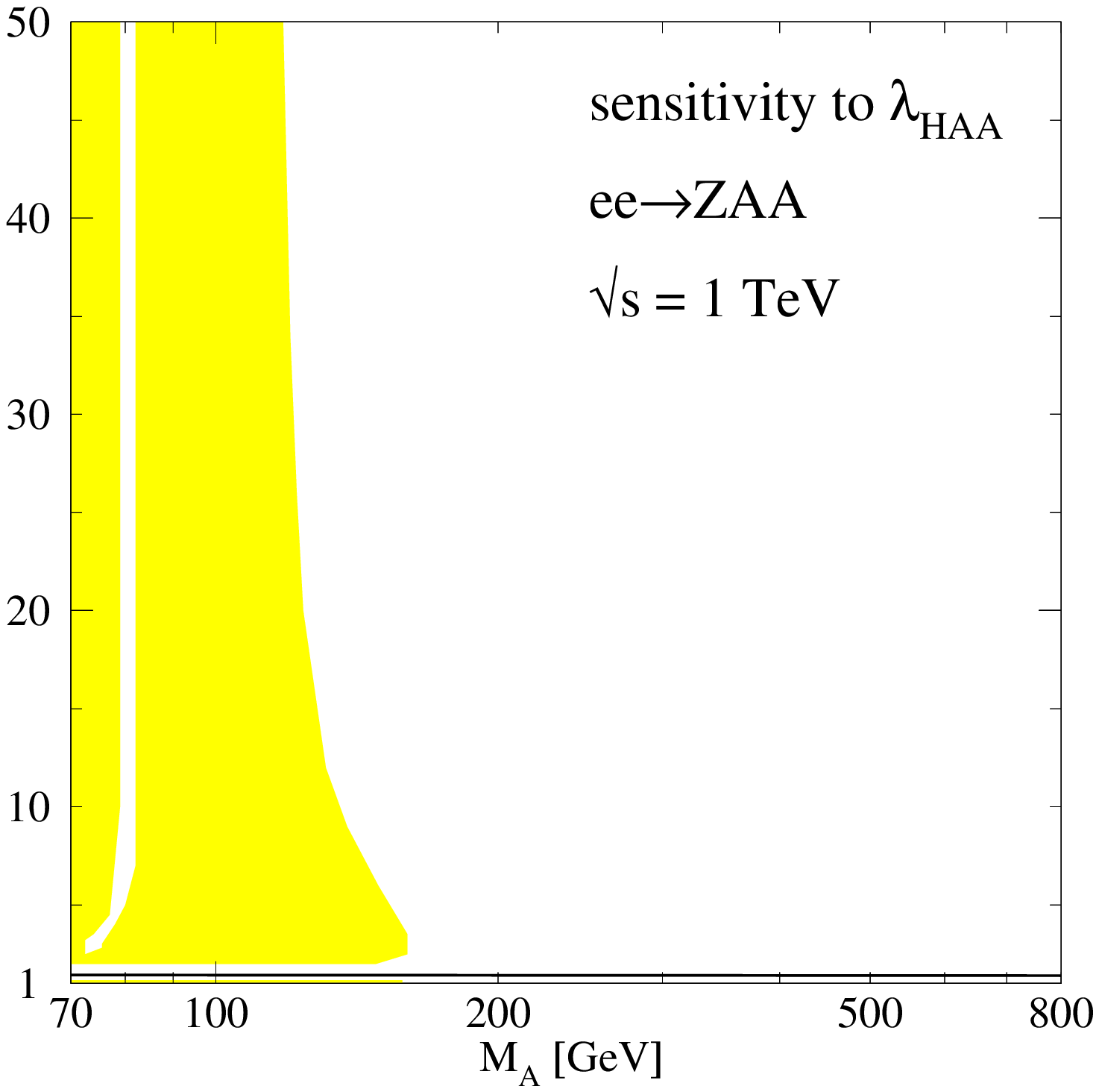,width=7cm}
\caption{Sensitivity to $\lambda_{hAA}$ and $\lambda_{HAA}$ in the 
process \ee$\to ZAA$ for a collider energy of $1$~TeV (no mixing).}
\label{fig:s3}
\end{center}
\end{figure}

Sensitivity areas of the trilinear couplings for the set of processes
defined in the correlation matrix Table~\ref{tab:coup}, are depicted
in Figs.~\ref{fig:s1} -- \ref{fig:s3}.  If at most one
heavy Higgs boson is present in the final states, the lower energy
$\sqrt{s}=500$~GeV is most preferable in the case of double
Higgs-strahlung. $HH$ final states in double Higgs-strahlung and
triple Higgs production involving $A$ give rise to larger sensitivity
areas at the high energy $\sqrt{s}=1$~TeV; increasing the energy to
1.6~TeV does not improve on the signal as a result of the scaling
behavior of the Higgs-strahlung cross section.  Apart from small
regions in which interference effects play a role, the magnitude of
the sensitivity regions in the parameter tan$\beta$ is readily
explained by the magnitude of the parameters $\lambda \sin
(\beta-\alpha)$ and $\lambda \cos (\beta-\alpha)$, shown individually
in Figs.~\ref{fig:lambda1} and \ref{fig:lambda2}. For large $M_A$ the
sensitivity criteria cannot be met any more either as a result of
phase space effects or due to the suppression of the $H$, $A$,
$H^\pm$ propagators for large masses. While the trilinear coupling of
the light neutral CP-even Higgs boson is accessible in nearly the
entire MSSM parameter space, the regions for $\lambda$'s involving
heavy Higgs bosons are rather restricted.\s

Since neither experimental efficiencies nor background related cuts
are considered in this paper, the areas shown in Figs.~\ref{fig:s1},
\ref{fig:s2} and \ref{fig:s3} must be interpreted as maximal
envelopes. They are expected to shrink when experimental efficiencies
are properly taken into account; more elaborate cuts on signal and
backgrounds, however, may help reduce their impact.\pskip

\subsection*{4. Conclusions}

In the present paper we have analyzed the production of Higgs boson
pairs and triple Higgs final states at $e^+ e^-$ linear colliders.
They will allow us to measure fundamental trilinear Higgs
self-couplings. The first theoretical steps into this area have been
taken by calculating the production cross sections in the Standard
Model for Higgs bosons in the intermediate mass range and for Higgs
bosons in the minimal supersymmetric extension.  Earlier results have
been combined with new calculations in this analysis. \s

The cross sections in the Standard Model for double Higgs-strahlung,
triple Higgs pro\-duc\-tion and double-Higgs fusion are small so that
high luminosities are needed to perform these experiments. Even though
the $e^+ e^-$ cross sections are below the hadronic cross sections,
the strongly reduced number of background events renders the search
for the Higgs-pair signal events, through $bbbb$ final states for
instance, easier in the $e^+ e^-$ environment than in jetty LHC final
states.  For sufficiently high luminosities even the first phase of
these colliders with an energy of 500 GeV will allow the experimental
analysis of self-couplings for Higgs bosons in the intermediate mass
range. \s

The extended Higgs spectrum in supersymmetric theories gives rise to a
plethora of trilinear and quadrilinear couplings. The $hhh$ coupling
is generally quite different from the Standard Model. It can be
measured in $hh$ continuum production at $e^+e^-$ linear colliders.
Other couplings between heavy and light MSSM Higgs bosons can be
measured as well, though only in restricted areas of the $[M_A,$
tan$\beta]$ parameter space as illustrated in the set of 
Figs.~\ref{fig:s1} -- \ref{fig:s3}. \pskip

\subsubsection*{Acknowledgements}

W.K.\ has been supported by the German Bundesministerium f\"ur Bildung
und Forschung (BMBF), Contract Nr.~05~6HD~91~P(0). We gratefully
acknowledge discussions with M.~Drees, M.~Du\-bi\-nin, H.~Haber,
P.~Lutz, L.~Maiani, P.~Osland, P.~Pandita F.~Richard and R.~Settles.
Special thanks go to M.~Spira for providing us with a program for
2-loop SUSY couplings, and for the careful reading of the manuscript.

\newpage

\newcommand{\sla}[1]{#1\!\!\!/}
\newcommand{\ewwhh}{\eta_{\ensuremath{\mbox{\tiny WWHH}}}}
\newcommand{\ewwaa}{\eta_{\ensuremath{\mbox{\tiny WWAA}}}}
\newcommand{\ewwH}{\eta_{\ensuremath{\mbox{\tiny WWH}}}}
\newcommand{\ehhh}{\eta_{\ensuremath{\mbox{\tiny HHH}}}}
\newcommand{\ewwh}{\eta_{\ensuremath{\mbox{\tiny WWh}}}}
\newcommand{\eHHh}{\eta_{\ensuremath{\mbox{\tiny HHh}}}}
\newcommand{\ehhH}{\eta_{\ensuremath{\mbox{\tiny hhH}}}}
\newcommand{\eHaa}{\eta_{\ensuremath{\mbox{\tiny HAA}}}}
\newcommand{\ehaa}{\eta_{\ensuremath{\mbox{\tiny hAA}}}}
\newcommand{\ewgh}{\eta_{\ensuremath{\mbox{\tiny $W^+G^-H$}}}}
\newcommand{\ewghh}{\eta_{\ensuremath{\mbox{\tiny $W^+G^-h$}}}}
\newcommand{\ewhmh}{\eta_{\ensuremath{\mbox{\tiny $W^+H^-H$}}}}
\newcommand{\ewhmhh}{\eta_{\ensuremath{\mbox{\tiny $W^+H^-h$}}}}
\newcommand{\ezah}{\eta_{\ensuremath{\mbox{\tiny ZAH}}}}
\newcommand{\ezahh}{\eta_{\ensuremath{\mbox{\tiny ZAh}}}}
\newcommand{\ewhma}{\eta_{\ensuremath{\mbox{\tiny $W^+H^-A$}}}}
\newcommand{\bw}{\beta_{\ensuremath{\mbox{\tiny W}}}}
\newcommand{\bh}{\beta_{\ensuremath{\mbox{\tiny H}}}}
\newcommand{\ba}{\beta_{\ensuremath{\mbox{\tiny A}}}}
\newcommand{\bp}{\beta_+}
\newcommand{\xh}{x_{\ensuremath{\mbox{\tiny H}}}}
\newcommand{\xhpm}{x_{\ensuremath{\mbox{\tiny $H^\pm$}}}}
\newcommand{\aw}{a_{\ensuremath{\mbox{\tiny W}}}}
\newcommand{\aH}{a_{\ensuremath{\mbox{\tiny H}}}}
\newcommand{\ah}{a_h}
\newcommand{\az}{a_{\ensuremath{\mbox{\tiny Z}}}}
\newcommand{\aA}{a_{\ensuremath{\mbox{\tiny A}}}}
\newcommand{\ahpm}{a_{\ensuremath{\mbox{\tiny $H^\pm$}}}}
\newcommand{\xn}{x_0}
\newcommand{\yn}{y_0}
\newcommand{\zn}{z_0}
\newcommand{\bb}{\beta}
\newcommand{\In}[1]{\ensuremath{\mbox{\tiny#1}}}
\newcommand{\lHh}{\lambda_{\ensuremath{\mbox{\tiny Hh}}}}
\newcommand{\mz}{M_{\ensuremath{\mbox{\tiny Z}}}}
\newcommand{\mw}{M_{\ensuremath{\mbox{\tiny W}}}}
\newcommand{\mH}{M_{\ensuremath{\mbox{\tiny H}}}}
\newcommand{\mh}{M_{\ensuremath{\mbox{\tiny h}}}}
\newcommand{\lij}{\lambda_{ij}}
\newcommand{\lia}{\lambda_{\ensuremath{\mbox{\tiny iA}}}}
\newcommand{\lhAA}{\lambda_{\ensuremath{\mbox{\tiny hAA}}}}
\newcommand{\lHAA}{\lambda_{\ensuremath{\mbox{\tiny HAA}}}}
\newcommand{\lhhphm}{\lambda_{\ensuremath{\mbox{\tiny $hH^+H^-$}}}}
\newcommand{\lHhphm}{\lambda_{\ensuremath{\mbox{\tiny $HH^+H^-$}}}}
\newcommand{\xp}{x_+}
\newcommand{\xw}{x_{\ensuremath{\mbox{\tiny W}}}}
\newcommand{\xa}{x_{\ensuremath{\mbox{\tiny A}}}}
\newcommand{\Xh}{x_{\ensuremath{\mbox{\tiny h}}}}
\newcommand{\XH}{x_{\ensuremath{\mbox{\tiny H}}}}
\newcommand{\sw}{\sin\theta_W}
\newcommand{\cw}{\cos\theta_W}
\newcommand{\sh}{\hat{s}}
\newcommand{\lh}{\lambda_{\ensuremath{\mbox{\tiny $hH_i H_j$}}}}
\newcommand{\lH}{\lambda_{\ensuremath{\mbox{\tiny $HH_i H_j$}}}}
\newcommand{\ct}{\cos\theta}
\newcommand{\ctt}{\cos^2 \theta}
\newcommand{\eep}{e^+ e^-}
\newcommand{\ra}{\rightarrow}



\subsection*{Appendix A: Double Higgs-strahlung Processes} 

\setcounter{equation}{0}
\renewcommand{\theequation}{A.\arabic{equation}}

In this appendix, we present the cross sections for the pair
production of the heavy MSSM Higgs bosons in the Higgs-strahlung
processes, $\eep \to Z H_i H_j $ and $ZAA$ with $H_{i,j}=h,H$. The
mixed process $\eep \ra Z H_iA $ is generated at the tree level by
gauge couplings only.  The notation is the same as in Section 2.1. The
trilinear couplings have been introduced earlier. Modifications of the
SM Higgs-gauge coupling in the MSSM are accounted for by the mixing
parameters:
\beq
\begin{array}{l@{:\;\;}l@{=\;}l l@{:\;\;}l@{=\;}l l@{:\;\;}l@{=\;}l}
VVh & d_1 & \sin(\beta-\alpha)\;\;\;\; & VVH & d_2 & 
\cos(\beta-\alpha) \;\;\;\;& VVA & d_3 & 0 \\
VAh & c_1 & \cos(\beta-\alpha) \;\;\;\;& VAH & c_2 & 
-\sin(\beta-\alpha) \;\;\;\;& WAH & c_3 & 1
\end{array}
\eeq
for $V=Z$ and $W$. The Higgs bosons are neutral except in the last 
entry.

\subsubsection*{A1: $\eep \to Z H_i H_j$} 

The double differential cross section of the process $e^+ e^- \to ZH_i
H_j$ is given for unpolarized beams by the expression
\beq
\frac{d\sigma[e^+ e^- \to ZH_i H_j]}{dx_1 dx_2} &=&  
\frac{ \sqrt{2} \, G_F^3 \, M_Z^6}{ 384\, \pi^3 s\,} 
\frac{ v_e^2+a_e^2}{(1-\mu_Z)^2} \  {\cal Z}_{ij} 
\label{eq:sigma}
\eeq
In terms of the variables $y_1, y_2, y_3$ defined in Section 2.1, and
with $\mu_i=M_{H_i}^2/s,$ $\mu_{ij}=\mu_i-\mu_j,$ etc., the 
coefficient ${\cal Z}_{ij}$ in the cross sections can be written as
\beq
{\cal Z}_{ij} &=& {\mathfrak a}^2_{ij}\,f_0 
+ \frac{{\mathfrak a}_{ij}}{2} \, \left[
  \frac{ d_i d_j \, f_3 }{y_1+\mu_{iZ}} 
+ \frac{ c_i c_j \, f_3}{y_1+\mu_{iA}} 
   \right] +\frac{ (d_i d_j)^2}{4\mu_Z (y_1+\mu_{iZ})}
\left[\frac{f_1}{y_1+\mu_{iZ}}+ \frac{f_2}{y_2+\mu_{jZ}} \right]\non \\
&+&  \!\!\!
\frac{ (c_i c_j )^2 }{4\mu_Z (y_1+\mu_{iA})}
\left[
\frac{f_1}{y_1+\mu_{iA}} 
+ \frac{f_2}{y_2+\mu_{jA}}\right] 
+ \frac{ d_i d_j c_i c_j} {2 \mu_Z (y_1+\mu_{iA})}
\left[
\frac{f_1}{y_1+\mu_{iZ}}
+\frac{f_2}{y_2+\mu_{jZ}} \right] 
\non\\
&+& \!\!\!
\Bigg\{ (y_1,\mu_i) \leftrightarrow (y_2,\mu_j) \Bigg\} 
\eeq
with 
\beq
{\mathfrak a}_{ij} = 
 \left[ \frac{ d_1 \lambda_{hH_iH_j}}{y_3 - \mu_{1Z}} + 
\frac{ d_2 \lambda_{HH_iH_j}}{y_3 - \mu_{2Z}} \right] + 
\frac{2 d_i d_j}{y_1+\mu_{iZ}} + \frac{2 d_i d_j}{y_2+\mu_{jZ}}
+ \frac{\delta_{ij}}{\mu_Z}
\eeq
The coefficients $f_{0}$ to $f_3$ are given by
\beq
f_0 &=& \mu_Z [(y_1+y_2)^2 + 8\mu_Z]/8 \non\\
f_1 &=& (y_1-1)^2(\mu_Z-y_1)^2-4\mu_iy_1(y_1+y_1\mu_Z-4\mu_Z) \non\\
& & + \mu_Z(\mu_Z-4\mu_i)(1-4\mu_i)-\mu_Z^2 + (\mu_i-\mu_j)^2 
[y_1(y_1-2)+1-4\mu_i] 
\non\\
& & + (\mu_i-\mu_j)
[8\mu_i(-y_1-\mu_Z)+2y_1 \mu_Z(y_1-2) + 2\mu_Z + 2y_1(y_1-1)^2]
\non\\
f_2 &=& [\mu_Z(1+\mu_Z - y_1 -y_2 - 8\mu_i)-(1+\mu_Z)y_1 y_2]
(2+2\mu_Z -y_1-y_2)
\non\\
& & + y_1 y_2[y_1 y_2 + \mu_Z^2+1+4\mu_i (1+\mu_Z)]
+ 4\mu_i \mu_Z(1+\mu_Z+4\mu_i)+ 
\mu_Z^2 \non\\
& & -2(\mu_i-\mu_j)^3-(\mu_i-\mu_j)^2[y_2(y_1-1)+10\mu_Z +4 \mu_j 
+3y_1 -1]\non\\
& & +(\mu_i-\mu_j)[\mu_Z(2(-y_1 y_2-y_1-8\mu_j)+6(\mu_Z+1-y_2))\non\\
& & +y_1((y_2-1)^2 -
y_1(1+y_2))+y_2(y_2-1)-4\mu_j(y_1-y_2)] \non\\
f_3 &=& y_1(y_1-1)(\mu_Z-y_1)-y_2(y_1+1)(y_1+\mu_Z)
+2\mu_Z(\mu_Z+1-4\mu_i) \non\\
& & + 2(\mu_i-\mu_j)^2 - (\mu_i-\mu_j)[y_2+y_1^2-3y_1+y_1y_2-4\mu_Z]
\eeq
For resonance contributions, propagator products have to be
substituted by $\pi_1(\mu_i) \pi_2(\mu_j) \to Re~\{\pi_1(\mu_i)
\pi_2(\mu_j^*) \}$ with $\mu_i \to \mu_i - i \gamma_i$ and $\gamma_i =
M_{H_i} \Gamma_{H_i}/s$.  Setting $i=j=1$, one recovers the
expressions eqs.~(\ref{zhh1}--\ref{zhh3}) for the process $\eep \to Zhh$.

\subsubsection*{A2: $\eep \to ZAA$} 

The differential cross section of the process $e^+ e^- \to ZAA$ 
is given by inserting ${\cal Z}_{33}$ in eq.~(\ref{eq:sigma}): 
\beq
{\cal Z}_{33} &=&  {\mathfrak a}^2_{33} \,  
f_0 + \frac{{\mathfrak a}_{33}}{2} \, \left[
\frac{c_1^2}{y_1-\mu_{1A}}+\frac{ c_2^2}{y_1-\mu_{2A}} 
\right]f_3  
\non \\
& & 
+ \frac{1}{4 \mu_Z} \left[ \frac{c_1^2} {y_1-\mu_{1A}}+ 
\frac{c_2^2}{y_1-\mu_{2A}} \right] \left[ 
\frac{c_1^2}{y_2-\mu_{1A}}+ 
\frac{c_2^2}{y_2-\mu_{2A}} \right]f_2  
\non \\
& &  
+ \frac{1}{4\mu_Z} \left[ 
\frac{c_1^2}{y_1-\mu_{1A}}+ 
\frac{c_2^2}{y_1-\mu_{2A}} 
\right]^2 f_1  
+ \Bigg\{ y_1 \leftrightarrow y_2 \Bigg\} 
\eeq
where 
\beq
{\mathfrak a}_{33} = 
\left[ \frac{ d_1 \lambda_{hAA}}{y_3 - \mu_{1Z}} + 
\frac{ d_2 \lambda_{HAA}}{y_3 - \mu_{2Z}} \right]
+  \frac{1}{\mu_Z} 
\eeq 
The coefficients $f_0$ to $f_3$ follow from the previous subsection 
after substituting $\mu_1,$ $\mu_2\to \mu_A$.



\subsection*{Appendix B: Triple Higgs Boson Production} 

\setcounter{equation}{0}
\renewcommand{\theequation}{B.\arabic{equation}}

The cross sections for the triple Higgs boson production of MSSM Higgs
bosons, $\eep \to A H_i H_j$ and $\eep \ra AAA$ with $H_{i,j}=h,H$ are
presented in this second appendix. The remaining process $\eep \to
H_iA A$ does not occur at tree level because of CP--invariance.

\subsubsection*{B1: $\eep \to A H_i H_j$} 

The double differential cross section of the process $e^+ e^- \to A
H_i H_j$ for unpolarized beams reads in the same notation as above:
\beq 
\frac{d\sigma}{dx_1 dx_2} = 
\frac{G_F^3 M_Z^6}{768 \sqrt{2} \pi^3 s} \, 
\frac{v_e^2+a_e^2}{(1-\mu_Z)^2} \, {\mathfrak A}_{ij} 
\eeq
with the function ${\mathfrak A}_{ij}$ 
\beq
{\mathfrak A}_{ij} &=&  \left[ 
\frac{ \lambda_{hH_iH_j} c_1} {y_3-\mu_{1A}}
+ \frac{ \lambda_{H H_iH_j} c_2} {y_3-\mu_{2A}} \right]^2 g_0
+ \frac{\lambda^2_{H_jAA} c_i^2} {(y_1+\mu_{iA})^2} g_1
+ \frac{ \lambda^2_{H_iAA} c_j^2} {(y_2+\mu_{jA})^2} g_1' \non \\
&& + \frac{c_j^2 d_i^2 } {(y_1+\mu_{iZ})^2} g_2 +
\frac{c_i^2 d_j^2 } {(y_2+\mu_{jZ})^2} g_2'+
\left[ \frac{\lambda_{hH_iH_j} c_1 } {y_3-\mu_{1A}}
+ \frac{ \lambda_{H H_iH_j} c_2 } {y_3-\mu_{2A}} \right] \non \\
&& \times 
\left[ \frac{ \lambda_{H_jAA} c_i} {y_1+\mu_{iA}} g_3
+ \frac{ \lambda_{H_iAA} c_j} {y_2+\mu_{jA}} g_3' 
+ \frac{c_j d_i } {y_1+\mu_{iZ}} g_4 +
\frac{c_i d_j } {y_2+\mu_{jZ}} g_4' \right] \non \\
&&+ 
\frac{\lambda_{H_i AA} \lambda_{H_j AA} c_ic_j}{(y_1+\mu_{iA})(y_2+\mu_{jA})}g_5
+ \frac{c_ic_j d_i d_j}{(y_1+\mu_{iZ})(y_2+\mu_{jZ})}g_8 
\non \\
&& + 
\frac{ \lambda_{H_j AA} c_ic_j d_i}{(y_1+\mu_{iA})(y_1+\mu_{iZ})}g_6
+ \frac{\lambda_{H_i AA} c_ic_j d_j}{(y_2+\mu_{jA})(y_2+\mu_{jZ})}
g_6' 
\non \\
&& 
+ \frac{\lambda_{H_j AA} c_i^2 d_j}{(y_1+\mu_{iA})(y_2+\mu_{jZ})} 
g_7 
+ \frac{\lambda_{H_i AA} c_j^2 d_i}{(y_2+\mu_{jA})(y_1+\mu_{iZ})}
g_7'
\eeq
The following expressions must be inserted for the coefficients $g_k$:
\beq
g_0 &=& \mu_Z[(y_1+y_2)^2-4\mu_A] \non \\
g_1 &=& \mu_Z(y_1^2-2y_1-4\mu_i+1) \non \\
g_2 &=& \mu_Z (2y_1 + y_1^2 -4y_2 +4y_2^2 + 4y_1 y_2 +1+ 4\mu_i 
- 8\mu_j - 8\mu_A) + (\mu_j-\mu_A)^2 \non\\
&& [8+(-2y_1+y_1^2-4\mu_i+1)/\mu_Z] + 2(\mu_j-\mu_A) 
(2y_1 y_2+y_1^2+2y_2-1)\non\\
g_3 &=& 2\mu_Z(y_1^2+y_1 y_2- y_1+y_2+2\mu_j-2\mu_i-2\mu_A) \non \\
g_4 &=& 2\mu_Z(y_1-y_2+y_1^2+2y_2^2+3y_1 y_2 -2\mu_j +2 \mu_i-6 
\mu_A) \non\\ 
&&+2(\mu_j-\mu_A)(-y_1+y_2+ y_1^2 + y_1y_2+2\mu_j-2\mu_A-2\mu_i) 
    \non \\
g_5 &=& 2\mu_Z(y_1+ y_2+y_1y_2+2\mu_j+ 2\mu_i-2\mu_A-1) \non \\
g_6 &=& 2\mu_Z( y_1^2+2 y_1y_2+2y_2+4\mu_j-4\mu_A -1) \non \\
 && + 2(\mu_j-\mu_A)(y_1^2-2y_1-4\mu_i+1) \non \\
g_7 &=& 2[ \mu_Z(2y_1^2+y_1y_2+y_2-3y_1+2\mu_j-6\mu_i-2\mu_A+1) 
\non \\
 && + (\mu_i-\mu_A)(y_1+y_2+y_1y_2+2\mu_j+2\mu_i-2\mu_A-1)] \non \\
g_8 &=& 2 \left\{\mu_Z(y_1+y_2+2y_1^2+2y_2^2+5y_1 y_2 -1 + 2\mu_j 
+ 2\mu_i -10\mu_A)  \right. \non \\
&& +2(\mu_i-\mu_A)(\mu_i-3\mu_j-\mu_A-2y_2+1) 
+2(\mu_j-\mu_A)(\mu_j-3\mu_i-\mu_A-2y_1+1)  \non \\
&& + [(\mu_j-\mu_A)((1+y_1)(y_2+2y_1-1)\mu_Z
+2\mu_i^2+4\mu_A^2 +\mu_A-\mu_i)\non \\
&& + (\mu_i-\mu_A) ((1+y_2) (y_1+2y_2-1)\mu_Z 
+ 2\mu_j^2+4\mu_A^2+\mu_A-\mu_j) \non \\
&& + \left. 6\mu_A(\mu_A^2-\mu_i\mu_j) + 
(\mu_i-\mu_A)(\mu_j-\mu_A) (1+y_1)(1+y_2)]/\mu_Z 
\right\}
\eeq
and 
\beq
g_k' (y_1,y_2,\mu_i,\mu_j) = g_k(y_2,y_1,\mu_j,\mu_i) 
\eeq



\subsection*{Appendix C: Heavy Higgs Production in $W_L W_L$ Fusion} 

\setcounter{equation}{0}
\renewcommand{\theequation}{C.\arabic{equation}}

In this third appendix we list the amplitudes and cross sections for the
pair production of the CP--even Higgs bosons in the longitudinal $W$
approximation $W_L W_L \ra H_i H_j$, as well as for $W_L W_L \ra AA$.
The notation is the same as in section 2.2.

\subsubsection*{C1: $W_L W_L \to H_i H_j$ } 

The amplitudes for the process $W_L W_L \to H_i H_j$ are given by: 
\beq
{\cal M}_{LL} &=& \frac{G_F \hat{s}}{\sqrt{2}} \left\{ 
(1+\beta_W^2)\left[\delta_{ij} + 
\frac{\lambda_{hH_iH_j} d_1}{(\hat{s} -M_h^2)/M_Z^2}
+ \frac{\lambda_{H H_iH_j} d_2}{(\hat{s}-M_H^2)/M_Z^2} \right] \right. 
\non \\
&&+ \frac{d_i d_j}{\beta_W \lambda_{ij}} \left[\frac{  r_W
+ (\beta_W -\lambda_{ij} \cos \theta )^2} {\cos \theta-x_W} -
\frac{ r_W + (\beta_W +\lambda_{ij} \cos \theta )^2}{\cos \theta+x_W} 
\right] \non \\
&&+ \left. \frac{c_i c_j}{\beta_W \lambda_{ij}} \left[ \frac{ 
r_+ + (\beta_W -\lambda_{ij} \cos \theta )^2 }{\cos \theta-x_+} - 
\frac{ r_+ + (\beta_W +\lambda_{ij} \cos \theta )^2 }{\cos \theta+x_+} 
\right] \right\}
\eeq
where $\mu_{i,j}= M_{H_{i,j}}^2/\hat{s}$,
$\beta_W=(1-4M_W^2/\hat{s})^{1/2}$ and $\lambda_{ij}$ is the usual
two--body phase space function, $\lambda^2_{ij} = (1-\mu_i -
\mu_j)^2-4\mu_i\mu_j$. Furthermore,
\beq
\begin{array}{l@{\quad\quad}l}
\hspace{-0.7cm}x_W= (1-\mu_i -\mu_j)/(\beta_W \lambda_{ij}) &
\hspace{-0.4cm}r_W= 1-\beta_W^4 -\beta^2_W (\mu_i-\mu_j)^2 \\[0.1cm]
\hspace{-0.7cm}x_+= (1-\mu_i -\mu_j+2 M_{H^\pm}^2/\hat{s} 
-2M_W^2/\hat{s})/(\beta_W\lambda_{ij}) &
\hspace{-0.4cm}r_+ =  -\beta^2_W (\mu_i-\mu_j)^2 
\end{array}
\eeq
The total cross section of the subprocess is obtained by integrating
the squared amplitude over the scattering angle; the result may be
written:
\beq
\sigma_{LL} (H_i H_j) &=& \frac{1}{1+\delta_{ij}} 
\frac{G_F^2 M_W^4}{2\pi \hat{s}}
\frac{\lambda_{ij}}{ \beta_W (1-\beta_W^2)^2} \non \\
&& \Bigg\{ (1+\beta_W^2)^2 \left[ \delta_{ij} 
+ \frac{\lambda_{hH_iH_j} d_1}{(\hat{s} -M_h^2)/M_Z^2} 
+ \frac{\lambda_{H H_iH_j} d_2}{(\hat{s} -M_H^2)/M_Z^2} \right]^2  
\non \\
&& +  \frac{2(1+\beta_W^2)} {\beta_W \lambda_{ij} }  \left[
\delta_{ij}  + \frac{\lambda_{hH_iH_j}d_1}{(\hat{s} -M_h^2)/M_Z^2}
 + \frac{\lambda_{H H_iH_j}d_2}{(\hat{s} -M_H^2)/M_Z^2} \right] \left[ 
d_i d_j a_1^W  + c_i c_j a_1^+   \right] \non \\
&&+ \left( \frac{d_i d_j}{\beta_W \lambda_{ij} } \right)^2 a_2^W  
+ \left( \frac{c_i c_j}{\beta_W \lambda_{ij} } \right)^2 a_2^+  + 4 
\left(\frac{d_i d_jc_i c_j}{\beta_W^2 \lambda^2_{ij} } \right) [a_3^W 
+ a_3^+ ] \Bigg\} 
\eeq
with 
\beq
a_1^W  &=& [ (x_W \lambda_{ij} -\beta_W)^2 + r_W ] \log \frac
{x_W-1}{x_W+1} +2 \lambda_{ij} (x_W \lambda_{ij} -2\beta_W ) \non \\
a_2^W &=& \left[ \frac{1}{x_W} \log \frac{x_W-1}{x_W+1} + \frac{2}
{x_W^2-1} \right] \bigg[ x_W^2 \lambda_{ij}^2 (3 \lambda_{ij}^2 x_W^2 
+2 r_W +14 \beta_W^2) \non \\
&&  -(\beta_W^2+ r_W)^2   -4 \lambda_{ij} \beta_W x_W 
 (3 \lambda_{ij}^2 x_W^2 +\beta_W^2+r_W ) \bigg] \non \\
&& - \frac{4}{x_W^2-1} \left[ \lambda_{ij}^2 (\lambda_{ij}^2 x_W^2 
+4 \beta^2_W - 4 \lambda_{ij} x_W \beta_W) - (\beta_W^2 +r_W)^2 
\right] \non \\
a_3^W&=& \frac{1}{x_+^2 - x_W^2} \, \log \frac{x_W-1}{x_W+1} \bigg[
2 \beta_W \lambda_{ij} x_W [(\beta_W^2+x_W^2 \lambda_{ij}^2)(x_W+x_+)
+ x_W r_W +x_+ r_+]  \non \\
&&  -x_+( r_+ + r_W + \lambda_{ij}^2 x_W^2)(\beta_W^2 + 
\lambda_{ij}^2 x_W^2)
-\beta_W^2( x_+ \beta_W^2 +4\lambda_{ij}^2 x_W^3+
x_+ x_W^2 \lambda_{ij}^2 ) \non \\
&&- x_+ r_W r_+  \bigg] + \lambda_{ij}^2 
\left[ \lambda_{ij}^2 x_+ x_W -2 
\beta_W \lambda_{ij} (x_W+x_+) + 4\beta_W^2   \right] \non 
\eeq
\beq
a_i ^+ \equiv a_i^W \ (x_W \leftrightarrow x_+ \ , 
\ r_W \leftrightarrow r_+ )
\eeq

\subsubsection*{C2: $W_L W_L \to AA $}

For pseudoscalar Higgs bosons, amplitude and cross section are
significantly simpler since only a few diagrams contribute to the $AA$
final state and, moreover, the masses of the final-state particles are
equal:
\beq
{\cal M}_{LL} &=& \frac{G_F \hat{s}}{\sqrt{2}}\left\{ 
(1+\beta_W^2) \left[1 + 
\frac{\lambda_{hAA} d_1} {(\hat{s} -M_h^2)/M_Z^2}
+ \frac{\lambda_{H AA} d_2} {(\hat{s} -M_H^2)/M_Z^2} \right] \right. 
\non \\
&&+ \left. \frac{1}{\beta_W \beta_A} \left[ 
\frac{ (\beta_W - \beta_A \cos \theta )^2}{\cos \theta-x_A} -  
\frac{ (\beta_W +\beta_A \cos\theta)^2}{\cos \theta+x_A} 
\right] \right\}
\eeq
with 
\beq
\beta_A=(1-4M_A^2/\hat{s})^{1/2} \quad \mathrm{and} \quad 
x_A=(1-2 M_A^2/\hat{s} +2 M_{H^\pm}^2/\hat{s} -2 M_W^2/\hat{s})
/(\beta_W \beta_A)
\eeq
The total cross section for the subprocess $W_L W_L \to AA$ may be cast 
into the form
\beq 
\!\!\!\!\!\!\!\!\!\!\!\!\sigma_{LL} (AA) 
\!\!&=&\!\! \frac{G_F^2 M_W^4}{4\pi\hat{s}}\frac{\beta_A}
{\beta_W (1-\beta_W^2)^2} \Bigg\{ (1+\beta_W^2)^2 \left[ 1 + 
\frac{\lambda_{hAA} d_1 } {(\hat{s} -M_h^2)/M_Z^2} + 
\frac{\lambda_{HAA} d_2 }{(\hat{s} -M_H^2)/M_Z^2} \right]^2  \non \\
&&+ 2(1+\beta_W^2) \left[ 1 + 
\frac{\lambda_{hAA} d_1 }{(\hat{s} -M_h^2)/M_Z^2}
\; + \; \frac{\lambda_{HAA} d_2 }{(\hat{s} -M_H^2)/M_Z^2} 
\; \right] \ \\
&& \times \frac{1}{\beta_W \beta_A}
\bigg[ (x_A \beta_A - \beta_W)^2 \log \frac{x_A-1}{x_A+1} 
 + 2\beta_A (x_A \beta_A -2\beta_W) \bigg] + \frac{1}{\beta_A^2 \beta_W^2} 
\times \non \\
&& \Bigg(
\log \frac{x_A-1}{x_A+1} \left[ 3 \beta_A^2 x_A (\beta_A x_A-2 \beta_W)^2
+\beta_W^2(2\beta_A^2 x_A -4 \beta_W \beta_A -\beta_W^2/x_A) \right] \non \\
&& 
+ \frac{2}{x_A^2-1} \left[ (3x_A^2 \beta_A^2-2\beta_A^2+\beta_W^2)
(\beta_Ax_A -2\beta_W)^2 + \beta_W^2 (\beta_A^2 x_A^2-3 \beta_W^2) 
\right] \Bigg) \Bigg\}  \non 
\eeq

\subsubsection*{C3: Asymptotia}

For asymptotic energies, the leading part of the $WW$ cross sections
does not involve the trilinear couplings $H_i H_j H_k$ or $H_i H_j A$.
Note however that the convoluted leptonic cross sections \ee
$\longrightarrow \bar{\nu}_e \nu_e H_i H_j$ and $AA$ are dominated by
the threshold regions, also for asymptotic \ee ~energies, so that the
observable leptonic high-energy cross sections are indeed sensitive in
leading order to the trilinear Higgs couplings.

\newpage


\end{document}